
\magnification=1200
\global\newcount\meqno
\def\eqn#1#2{\xdef#1{(\secsym\the\meqno)}
\global\advance\meqno by1$$#2\eqno#1$$}
%
\global\newcount\refno
\def\ref#1{\xdef#1{[\the\refno]}
\global\advance\refno by1#1}
\global\refno = 1
\vsize=7.5in
\hsize=5.8in
\tolerance 10000
%
%

\def\bx{\partial^2}

\def\calz{\cal Z}
\def\calzl{{\cal Z}_{k,\ell}}
\def\calzt{{\cal Z}_{k,t}}
\def\delab{\delta^{ab}}

\def\dell{\Delta_{\ell}}
\def\dellk{\Delta_{k,\ell}}
\def\delt{\Delta_t}
\def\deltk{\Delta_{k,t}}

\def\dta{\Delta}
\def\half{{1\over2}}

\def\item{\par\hang\textindent}

\def\km{K_0^{-1}}

\def\mat{(1+K_0^{-1}\delta K_0)^{-1}}
\def\matt{(1+K_0^{-1}\delta K_0)^{-2}}
\def\mattt{(1+K_0^{-1}\delta K_0)^{-3}}
\def\parmu{\partial_{\mu}}
\def\parnu{\partial_{\nu}}
\def\pmu{p_{\mu}}
\def\pnu{p_{\nu}}

\def\tphi{\tilde\phi}
\def\tphia{{\tilde\phi}^a}
\def\tphib{{\tilde\phi}^b}
\def\tphic{{\tilde\phi}^c}

\def\tc{\tilde {\cal Z}}

\def\tz{\tilde Z}

\def\tzt{{\tilde Z}_t}
\def\udot{ U'_k}
\def\uddot{U''_k}
\def\udddot{ U'''_k}
\def\uell{u_{\ell}}
\def\uellk{U^{(11)}_k}
\def\upp{U^{(11)}_{k} }

\def\ut{u_t}
\def\utk{U^{(22)}_k}
\def\utt{ U^{(22)}_{k} }

\def\za{Z^{(a)}}
\def\zab{Z^{(ab)}}
\def\zac{Z^{(ac)}}
\def\zb{Z^{(b)}}
\def\zbc{Z^{(bc)}}
\def\zc{Z^{(c)}}
\def\zcd{Z^{(cd)}}
\def\zell{Z_{\ell}}

\def\zellk{{\cal Z}_{k,\ell}}
\def\zt{Z_t}

%
%

\baselineskip 12pt plus 1pt minus 1pt
\vskip 2in
\centerline{{\bf RENORMALIZATION GROUP AND UNIVERSALITY}
\footnote{*}{This work is
supported in part by funds
provided by the U. S. Department of Energy (D.O.E.) under contract
\#DE-FG05-90ER40592.}}
\vskip 24pt
\centerline{Sen-Ben Liao and Janos Polonyi
\footnote{$^\dagger$}{
On leave from CRIP and R. E\"otv\"os University, Budapest, Hungary}}
\vskip 12pt
\centerline{\it Department of Physics}
\centerline{\it Duke University}
\centerline{\it Durham, North Carolina\ \ 27708\ \ \ U.S.A.}
\vskip 12pt
\centerline{and}
\vskip 12pt
\centerline{\it Laboratory of Theoretical Physics}
\centerline{\it and Department of Physics}
\centerline{\it Louis Pasteur University}
\centerline{\it 67087\ \ Strasbourg\ \ Cedex\ \ France}
\vskip 1in
\centerline{To appear in  {\it Phys. Rev. D}}
\vskip 0.7in
\vskip 24 pt
\baselineskip 12pt plus 2pt minus 2pt
\centerline{{\bf ABSTRACT}}
\medskip
It is argued that universality is severely limited for models with
multiple fixed points. As a demonstration
the renormalization group equations are presented for the potential and
the wave function renormalization
constants in the $O(N)$ scalar field theory. Our equations
are superior compared with the usual approach which retains only
the contributions that are nonvanishing in the ultraviolet regime.
We find an indication for the existence of relevant operators at the
infrared fixed point, contrary to common expectations. This result
makes the sufficiency of using only renormalizable coupling constants
in parametrizing the long distance phenomena questionable.
\vskip 24pt
\vfill
\noindent Duke-TH-94-64, LPT 94-3\hfill January 1995
\eject
\medskip
\baselineskip 12pt plus 2pt minus 2pt
\centerline{\bf I. INTRODUCTION}
\medskip
\nobreak
\xdef\secsym{1.}\global\meqno = 1
\medskip
The modification of
fundamental laws of physics with the change of observational length scale
is the subject of the renormalization group (RG) \ref\wilsko. Through the
RG flow equation one may probe the dependence
of the effective coupling constants on the characteristic length.
The otherwise complicated flow pattern
becomes rather simple in the vicinity of the fixed points where
the linearized RG flow along with scaling
provide a recipe for classifying the coupling constants via their
dependence on the characteristic scale.
The irrelevant coupling constants are those which decrease as the scale
is moved toward the infrared (IR) direction. The physical content of the
theory is insensitive
to the actual choice of these coupling constants in the IR end
of the region where the linearization of the RG
equation is applicable. Within this regime where the
usual concept of universality is recovered, the physics is parameterized
by the others only, namely, the relevant and marginal coupling constants.

In the realistic models we find several scaling regimes when the
renormalized trajectory passes by different fixed points. There
are fixed points for the Theory of Everything, GUT, Standard model,
QCD and QED, to mention some of them. Although the true renormalized
trajectory approaches all of them for certain values of the cutoff,
it reaches the first one only. In fact,
in the scaling regime of, say the fixed point of QCD, some of the interactions
of the Standard Model generate nonrenormalizable vertices
in terms of the quark and gluon fields \ref\decoupl. These vertices
deflect the
renormalized trajectory from the fixed point as we move up in energy.
(Hence it is physically not too crucial whether or not the ultraviolet (UV)
fixed point
really exists. All we shall assume here is the scaling up to a certain
energy scale.) The traditional goal of local field theory is to
give an account of these vertices in terms of elementary particle
exchanges in a manner which is renormalizable at higher energy.
Yet renormalization, i.e., the removal of the UV cutoff is
necessary only for the Theory of Everything. In fact, when the scaling
is investigated at the other fixed points then the higher energy
reactions always make these fixed points unstable
in the UV direction.

The scenario sketched above leads to a serious limitation of the
use of the concept of universality. It is true that there
are ``islands'' of autonomous scaling regimes where physics can be
parameterized
by relevant or marginal operators only, but these operators usually vary
with scaling regimes and the matching of
relevant coupling constants is rather nontrivial. Even though we can
establish the
the importance of certain coupling constants in a given energy range,
the physics at a different scale will be governed by
different set of coupling constants \ref\paris.

There is one last fixed point as we move towards longer distance scale,
the infrared (IR) fixed point. Macroscopic physics is characterized by
the scaling at this IR fixed point.
Can this fixed point have relevant operators ? The answer is
negative for theories with a mass gap. To see this it is sufficient to
recall that the dependence of the coupling constants on the cutoff is
to take into account the effects of the modes which are eliminated
as the cutoff energy is lowered. At energy well below the mass gap
the fluctuations are suppressed and the evolution of the
coupling constants slows down. Thus the IR limit of the theory is
stable.

The situation is more interesting for theories without mass gap.
Realistic theories with spontaneous symmetry breaking
belong to this class. For such theories, IR divergences can pile
up and generate relevant coupling constant in the IR regime.
It is the main result of this paper that this indeed happens in
simple four dimensional scalar models. In this case the long distance
physics of the model is not universal, i.e., it cannot be parameterized
completely by the relevant and marginal coupling constants of the
UV fixed point.

In arriving at this result one needs an improved version of the
RG equations. The usual method of renormalizing the
theory is to follow the mixing and the evolution of selectively few coupling
constants. One traditionally chooses those coupling constants which are
relevant or marginal in the vicinity of the UV fixed
point. As we lower the cutoff to the natural mass scale $m_R$
of the theory, the scaling properties change fundamentally. In the
IR side of the natural mass scale where $k < m_R$,
one deduces the scaling laws corresponding to the
IR fixed point. However, the UV and the IR scaling
regimes are separated by a crossover at $m_R$, and
there is no reason whatsoever to expect the same set of
relevant or marginal coupling operators for both UV and IR
fixed points.
Furthermore we do not have a simple power counting argument to find out
the relevant operators of the IR fixed point. Thus in order
to establish the scaling operators of the IR fixed point,
we have to trace down the evolution and the mixing of many more
operators which might well be irrelevant in the UV scaling regime.
This can be achieved with a RG flow
equation which is capable of handling the mixing between infinitely many
coupling constants.
Such an improved RG equation has
been obtained in \ref\wegner\ and was subsequently applied for the UV
scaling regime in \ref\wehar\ in the leading-order approximation of the
derivative
expansion for the renormalized action. We present in this paper
the RG equation \wegner\ applied in
the next order of the derivative expansion. This allows us to verify
our claim about the existence of the relevant operators at the
IR fixed point in the first two orders of the derivative expansion.

The organization of the paper is the following: In Sec. II we give a
brief derivation of the RG equation \wegner\
in the leading order of the derivative expansion for $U_k(\Phi)$ in
the one-component scalar model and show the emergence
of the IR singularities in certain $\beta$ functions. Section III contains
the technical details for deriving
$U_k(\Phi)$ for the $O(N)$ $\lambda\phi^4$ theory with two distinct
wave function renormalization constants, $\tilde\calzl$ and
$\tilde\calzt$ for the longitudinal
and the transverse components, respectively.
In Sec. IV, we derive a set of three
coupled nonlinear RG flow equations for $U_k(\Phi)$, $\tilde\calzl$
and $\tilde\calzt$. Asymptotic scalings in both UV and IR
limits are discussed in Sec. V. Section VI contains our conclusions.
Two appendices are included supplementing the detail of deriving
the RG equations for the paper.
\bigskip
\centerline{\bf II. ONE-COMPONENT SCALAR FIELD THEORY}
\medskip
\nobreak
\xdef\secsym{2.}\global\meqno = 1
\medskip
Our starting point for deriving the RG equation
for a system characterized by the field
$\phi(x)$ is to introduce the coarse-grained ``block variable'':
\eqn\kfiss{\phi_k(x)=\int_y\rho_k(x-y)\phi(y),}
where
\eqn\spint{\int_x=\int d^dx=\Omega ,}
via a smearing function $\rho_k(x)$, with $k^{-1}$ being the
characteristic linear dimension of the region over which
the field averaging is performed. In this paper, we shall choose
$\rho_k(x)$ to be
\eqn\shcr{\rho_k(x)=\int_{p<k} {d^dp\over (2\pi)^d}e^{ipx},}
or $\rho_k(p)=\Theta(k-p)$, i.e., a sharp cutoff \ref\lp.
Although $\rho_k(x)$ acts as an upper cutoff, we shall
use $\Lambda$ as the $k$-independent UV cutoff for the
theory.

Given a set of blocked variables $\phi_k(x)$,
the blocked action $\tilde S_k$ can be deduced from
\eqn\cactt{ e^{-\tilde S_k(\Phi)}=\int D[\phi]\prod_x
\delta(\phi_k(x)-\Phi(x))e^{-S(\phi)},}
where the field average $\Phi$ of a given block is chosen to coincide with
the slowly varying background. By performing the functional integration in
loop expansion subject to the $\delta$ function constraints one finds
\eqn\efftw{\eqalign{\tilde S_k(\Phi)&=S(\Phi) +{1\over 2}{\rm Tr'}
{\rm ln}{{\bx S}\over \partial\phi(p)\partial\phi(-p)}
-{1\over 2}\int_p^{'} {{\partial S}\over \partial\phi(p)}
\Biggl({{\bx S}\over \partial\phi(p)\partial\phi(-p)}\Biggr)^{-1}
{{\partial S}\over \partial\phi(-p)} \cr
&
=S(\Phi)+{1\over 2}{\rm Tr'}{\rm ln}K - {1\over 2}\int_p^{'}FK^{-1}F ,}}
\eqn\spi{\int_p^{'}=\int_k^{\Lambda}{d^dp\over (2\pi)^d},}
where $F$ and $K$ are the first and the second functional derivative of the
bare lagrangian, respectively,
and ${\rm Tr'}$ denotes the trace sum over internal space as well as the
restricted
momentum space with $k <  p < \Lambda$. How
blocking transformation modifies the propagator $\dta(x-y)=K^{-1}(x-y)$
can be seen explicitly by considering a free scalar theory:
\eqn\modi{ \int_p^{'} F(p)\dta(p)F(-p)
\longrightarrow \int_p F(p)\tilde\dta(p)F(-p),}
where
\eqn\fret{ \tilde\dta(x-y)=\int_p
{e^{ipx}\over {p^2+\mu^2}}\Theta(p-k) }
is the ``blocked'' propagator with an effective IR cutoff scale $k$.
In the limit $k\to 0$, one recovers the original $\Delta(x-y)$.

Eq. \efftw\ is far too complicated so the derivative expansion \ref\fraser\
is used at this point. The form
\eqn\blact{\tilde S_k(\Phi)=\sum_{n=0}^\infty\int d^dxL_k^{(n)}(\Phi(x))}
is assumed where $L_k^{(n)}(\Phi(x))$ is a homogeneous polynomial of
order $2n$ in the space-time derivatives. We shall truncate the
expansion at $n=1$, retaining only the wave function renormalization
function $Z_k(\Phi)$ and the blocked potential $U_k(\Phi)$.
Such truncation, being justifiable in the IR limit for high enough
space-time dimension $d$, yields simpler differential equations when
substituted into \efftw. In principle, however,
equations that generate the scale dependence of $Z_k(\Phi)$ and the
higher order derivative terms must also be calculated in the framework
of the derivative expansion in order to have a closed system.

It is worth mentioning that \efftw\ gives the one-loop effective
potential for $k=0$ \lp. The modes with nonvanishing wave number
are eliminated independently in the one-loop approximation. However, the
result can be greatly improved by the successive elimination of the degrees
of freedom. In the improved scheme, the contribution of a particular
mode which has been integrated out is kept for the elimination
of the next mode, thereby taking into account
the interactions between the modes. By decreasing the cutoff
infinitesimally from $k\to k-\Delta k$. we generate from \efftw\ the evolution
of the potential, $L^0_k(\Phi)=U_k(\Phi)$, which
for $\Phi(x)=\Phi$ and $d=4$, becomes:
\eqn\aprhw{U_{k-\Delta k}(\Phi)=U_k(\Phi)+\Delta k{k^3\over16\pi^2}
{\rm ln}\Bigl[{{Z_k(\Phi)k^2+\partial^2_\Phi U_k(\Phi)}\over
{Z_k(0)k^2+\partial^2_\Phi U_k(0)}}\Bigr],}
or equivalently, a differential equation of the form:
\eqn\aprdf{k{\partial_k}U_k(\Phi)=-{k^4\over16\pi^2}{\rm ln}
\Bigl[{{Z_k(\Phi)k^2+\partial^2_\Phi U_k(\Phi)}\over
{Z_k(0)k^2+\partial^2_\Phi U_k(0)}}\Bigr]}
in the limit $\Delta k\to0$.
This equation describes the renormalization of the potential
with arbitrary dependence on the field $\Phi$. The solution
for $U_{k=0}(\Phi)$ differs from the usual one-loop effective potential
mentioned
before insofar that the effects of the operators which are irrelevant
at the UV fixed point are retained during the elimination of the
degrees of freedom. This difference is negligible for a weakly coupled
theory as long as no new relevant operators are generated
outside the UV scaling regime.

Equation \aprdf\ can be derived by resumming the one-loop contributions to
evolution of the potential.  In fact, it can be written as
\eqn\aprdfp{\eqalign{k{\partial_k}U_k(\Phi)
&=-{k^4\over16\pi^2}\biggl\{ {\rm ln}\Bigl[{{Z_k(\Phi)k^2+m_k^2}\over
{Z_k(0)k^2+m_k^2}}\Bigr]+ {\rm ln}\Bigl[1+(Z_k(\Phi)k^2+m_k^2)^{-1}
\partial^2_\Phi V_k(\Phi)\Bigr]\biggr\}\cr
&=-{k^4\over16\pi^2}\biggl\{{\rm ln}\Bigl[{{Z_k(\Phi)k^2+m_k^2}\over
{Z_k(0)k^2+m_k^2}}\Bigr]+
{\partial^2_\Phi V_k(\Phi)\over Z_k(\Phi)k^2+m^2}-{1\over2}
\biggl({\partial^2_\Phi V_k(\Phi)\over Z_k(\Phi)k^2+m_k^2}\biggr)^2+\cdots
\biggr\},}}
where $\partial_\Phi^2 V_k(\Phi)=\partial_\Phi^2 U_k(\Phi)-m_k^2$
and $m_k^2=\partial_\Phi^2U_k(0)$. The last
line contains the sum of the one-loop graphs with increasing number of
$\partial^2_\Phi V_k(\Phi)$ insertions.
The external legs of these graphs which are attached to $\Phi$ in
$\partial_\Phi^2V_k(\Phi)$ are carrying
zero momentum and the modes with momentum $k$ are propagating along the loop.
This is just the set of graphs one has to sum up in eliminating the modes
with momentum $k$.

There are certainly higher loop corrections to \aprdfp. However, the
terms of the order $m$ in loops contains $m$ integrations over a
$d$-dimensional shell in the
momentum space. When only few modes are eliminated,
$\Delta k\approx0$, the integration over each shell
yields, on the dimensional ground, a new small parameter:
\eqn\eps{\zeta={\Delta k\over k},}
which helps suppress the higher loop contributions to the RG evolution equation
in \aprdf. This is the basis of the ``exactness''
of for a RG equation which is formally obtained in the one-loop
approximation.

As the IR limit is approached with $k\to0$, $\zeta$ becomes ill-defined. To
examine the behavior of $\zeta$ in this regime, we introduce an IR cutoff
by considering the system in a box of size $L$. With the number of
degrees of freedom being $N^d=({L\Lambda\over2\pi})^d$, the momentum
integral measure takes on the form:
\eqn\minm{{1\over N^d}\sum_{k_\mu}\longrightarrow\int_{-\pi}^\pi
{d^dp\over(2\pi)^d},}
in term of the dimensionless momentum, $p_\mu$. In the ``ultimate
RG transformation'', where only a single mode
is eliminated each time, the small parameter would be ${1\over N^d}$,
the inverse of the
number of degrees of freedom remained. Therefore,
the IR limit can be reached in two
different manners: (i) The limit $L\to\infty$ is taken first before
$k\to 0$; and  (ii) $k\to{2\pi\over L}$ is first taken for
finite systems followed by $L\to\infty$. We immediately notice that
$\zeta$ can be kept small only for case (i) and becomes
O(1) for case (ii). In another words, the limits $k\to0$ and
$L\to\infty$ do not necessarily commute. In fact,
the gap of the two dimensional $\sigma$-model, whose existence is
established with reasonable accuracy using procedure (ii) is absent if
(i) is employed instead \ref\patr.

It seems that if the two limits $k\to0$ and
$L\to\infty$ are not commuting, procedure (ii) would be more reasonable
to describe the dynamics of local interactions. In that case the
RG equation, \aprdf, can only represent
a partial resummation of the perturbation expansion. Such a loss
of the effectivity is due to the presence of the length scales,
$L$ and $k^{-1}$, in a system without IR mass gap. Since in the subsequent
treatments we implement the one-loop RG equation for the
IR regime, the conclusions drawn from such computation
are strictly relevant only for case (i). It remains to be seen if they
can be carried over to the procedure (ii).

Adopting (i) as our approach, we find indication of the emergence of
relevant operators at the IR
fixed point for massless theories due to the following handwaving argument:
For the sake of simplicity we keep
the wave function renormalization constant, $Z_k(\Phi)$, to be unity
in \aprdf\ and introduce the $\beta$ functions for the
coupling constants for $\Phi^n$ as:
\eqn\betaf{\beta_n(k)=\partial^n_\Phi k\partial_kU_k(\Phi),}
which can be obtained by substituting \aprdf\ into \betaf.
These $\beta$ functions describe the evolution of the coupling strengths of
small fluctuations around the constant background $\Phi(x)=\Phi$.
Let us now consider a model where
\eqn\masslcn{\partial^2_\Phi U_{k=0}(0)=0.}
This is what one commonly calls a ``massless'' theory
since $U_{k=0}(\Phi)$ is just the effective potential. However,
this name is misleading when spontaneous symmetry breaking occurs with
$<\phi(x)>\not=0$ since $\partial^2_\Phi U_{k=0}(<\phi(x)>)$ is now
non-vanishing. The
leading IR contribution for the \betaf\ comes from the highest
power of $k^2+\partial^2_\Phi U_k(0)$ in the denominator:
\eqn\betaapr{\beta_n(k)=(-1)^n{k^4\over16\pi^2}
\Biggl({\partial^3_\Phi U_k(0)\over{k^2+\partial^2_\Phi U_k(0)}}\Biggr)^n
(1+O(k^2)),}
which shows that for $\partial^3_\Phi U_k(0)>0$ the coupling constants for
the odd powers of the field blow up in the IR limit.
Although we have not found the scaling operators there ought
to be relevant ones which drive the IR divergences.

Such conclusion could have been reached by considering the graphs which
contribute to the evolution equation:
\eqn\aprdfp{\beta_n(k)=-{k^4\over16\pi^2}\partial^n_\Phi\biggl\{
{\partial^2_\Phi V_k(\Phi)\over Z_k(\Phi)k^2+m^2}-{1\over2}
\biggl({\partial^2_\Phi V_k(\Phi)\over Z_k(\Phi)k^2+m_k^2}\biggr)^2+\cdots
\biggr\}.}
In the UV scaling regime where $k^2>>\partial_\Phi^2U_k(\Phi)$,
the dominant contribution comes from the graphs with least number of
propagators. The evolution of the vertex with $n$ legs is described by joining
two legs of the $(n+2)$th order vertex having momentum $k$. One
reproduces the usual $\beta$ functions, e.g,:
\eqn\gammaus{\beta_2(k)=-{\partial_\Phi^4 U_k(\Phi)\over16\pi^2}
{k^4\over k^2+m_k^2},}
for the mass squared, and
in the usual one-loop approximation to the $\phi^4$ model where there is no
sixth-order vertex at the cutoff,
\eqn\betaus{\beta_4(k)={3\bigl(\partial_\Phi^4 U_k(\Phi)\bigr)^2\over16\pi^2}
{k^4\over(k^2+m_k^2)^2},}
in the next-to-leading-order approximation in \aprdfp.

As we enter the IR regime with $k^2<<\partial_\Phi^2U_k(\Phi)$,
the scaling laws quickly change. For a theory with mass gap the contributions
are $O({k^4\over m^4})$ and the evolution slows down indicating the
absence of relevant operators. But for massless theories,
${\rm lim}_{k\to0}m^2_k=0$,
the dominant contributions are received from graphs with the maximal
number of propagators between the vertices. In the absence of other
dimensional parameter, we find $\partial^2_\Phi V_k(0)\sim k^2$ and the IR
contribution to the evolution of the $n$-th order vertex is dominated
by the one-loop graph with $n$
insertion of the vertex $\partial_\Phi^3U_k(\Phi)$ as shown in \betaapr.

Unfortunately this result is not interesting. The theory develops a
nonvanishing vacuum expectation value
for the field either due to the masslessness
or the presence of the odd powers of the field in the potential. Thus, the
coupling constants computed at $\Phi=0$ are not characterizing the
strength of the interactions of small fluctuations in the vacuum. The true
vacuum with $<\Phi(x)>\not=0$ shields the IR divergences.
However, if the theory possesses a
continuous symmetry which is broken spontaneously then
Goldstone's theorem guarantees the presence of the massless modes in the
vacuum. The more careful repetition of this simple argument for
the $N$-component scalar field theory is the subject of this work.
\bigskip
\centerline{\bf III. DERIVATIVE EXPANSION FOR THE O(N) MODEL}
\medskip
\nobreak
\xdef\secsym{3.}\global\meqno = 1
\medskip
We consider a generalized bare $O(N)$ scalar field lagrangian of the form
\eqn\barelgr{{\cal L}(\phi)=\half\tilde Z_a(\parmu\phi^a)^2+V(\phi),}
with $a=1,\cdots,N$.
The theory is chosen to be in the symmetry broken phase and the
the direction $a=1$ is chosen to be in the expectation value of the field.
The extra subscript in $\tilde Z$ is used differentiate between
two wavefunction renormalization
constants, one for the longitudinal component and the other for the
transverse ones:
\eqn\ztv{\tilde Z_a=\cases{\tilde Z_{\ell},&$a=1$ \cr
\cr
\tilde Z_t, &$a=2,\cdots,N$.  \cr }}
Alternatively, one may write \barelgr\ as
\eqn\lgrr{\eqalign{{\cal L}(\phi)&={1\over 2}\tilde\zell(\parmu\phi^1)^2
+{1\over 2}\tilde\zt(\parmu\phi^i)^2+V(\phi) \cr
&
=-\half Z_a\phi^a\bx\phi^a+V(\phi)}}
via the relation:
\eqn\wfrcs{\cases{\eqalign{ \tilde\zell &={d\over d\phi^1}(\zell\phi^1)\cr
\tilde Z_t&={d\over d\phi^i}(Z_t\phi^i),~~~~ i=2,\cdots,N. \cr}}}
We shall split $\phi(x)$ into the slowly varying background $\chi(x)$,
and the fast-fluctuating modes $\xi(x)$ such that
\eqn\field{\phi^a(p)=\cases{\chi^a(p),&$0 \le p \le k$ \cr
\cr
\xi^a(p), &$k < p < \Lambda$.  \cr }}
Noting that $\phi_k(p)=\rho_k(p)\phi(p)=\chi(p)$, one then integrates out the
fast-fluctuating modes $\xi(x)$ by using the loop expansion
to obtain the blocked action $\tilde S_k(\Phi)$ as a function of the
blocked field average $\Phi=(\Phi^2)^{1/2}$:
\eqn\efact{\eqalign{\tilde S_k(\Phi)&=-{\rm ln}\int D[\chi]
D[\xi]\prod_x\delta(\phi_k(x)-\Phi(x))~{\rm exp}\bigl\{-S(\chi+\xi)\bigr\} \cr
&
=-{\rm ln}\int D[\chi]\prod_p\delta(\chi(p)-\Phi(p))\cr
\qquad\qquad
&\times\int D[\xi]~{\rm exp}\Bigl\{-S(\chi)-\int_p\xi^a(p)F^a(-p)
-{1\over 2}\int_p\xi^a(p)K^{ab}(p,-p)\xi^b(-p)+\cdots\Bigr\} \cr
&
=-{\rm ln}\int D[\chi]\prod_p\delta(\chi(p)-\Phi(p))~{\rm exp}
\Bigl\{-S(\chi)-{1\over 2}{\rm Tr'}{\rm ln}K +{1\over 2}
\int_p^{'}F K^{-1}F+\cdots\Bigr\} \cr
&
=S(\Phi) + {1\over 2}{\rm Tr'}~
{\rm ln}~ K(\Phi) - {1\over 2}\int_p^{'}FK^{-1}F\Big\vert_{\Phi} \cr
&
=S(\Phi)+\delta {\tilde S}^1_k(\Phi) + \delta {\tilde S}^2_k(\Phi) ,}}
where
\eqn\fkern{F^{a}(\Phi) = {{\partial S}
\over \partial\phi^a(x)}\Big\vert_{\Phi}
=-{1\over 2}Z_c^{(a)}(\Phi)\Phi^c\bx\Phi^c-{1\over 2}Z_a(\Phi)\bigl(
\bx\Phi^a+\Phi^a\bx\bigr)+ V^{(a)}(\Phi),}

\eqn\kerd{\eqalign{K^{ab}(\Phi) &= {{\partial^2 S}
\over \partial\phi^a(x)\partial\phi^b(y)}\Big\vert_{\Phi} \cr
&
=\Biggl\{-{1\over 2}Z_c^{(ab)}(\Phi)\Phi^c\bx\Phi^c-Z_a(\Phi)\delab\bx
+ V^{(ab)}(\Phi) \cr
&
-{1\over 2}Z_b^{(a)}(\Phi)(\bx\Phi^b+\Phi^b\bx)-{1\over 2}Z_a^{(b)}(\Phi)
(\bx\Phi^a+\Phi^a\bx)\Biggr\}\delta^4(x-y) ,}}
and
\eqn\zpr{\cases{\eqalign{ Z^{(a_1a_2\cdots a_n)}(\Phi)&={ {\partial^n Z}
\over {\partial\phi^{a_1}\partial\phi^{a_2}\cdots
\partial\phi^{a_n} }}\Big\vert_{\Phi} \cr
V^{(a_1a_2\cdots a_n)}(\Phi)&={{\partial^n V}
\over {\partial\phi^{a_1}\partial\phi^{a_2}\cdots
\partial\phi^{a_n} }}\Big\vert_{\Phi} .\cr}}}
Note that $\rm Tr'$ denotes the trace sum over the
internal symmetry space as well as the restricted space-time.

As noted in the Introduction the blocked action can be expanded as
\eqn\efactt{\tilde S_k(\Phi)=\int_x\Biggl(-{{\cal Z}_{a,k}(\Phi)\over2}
\Phi^a\bx\Phi^a + U_k(\Phi)+O(\partial^4)\Biggr).}
For the computation of ${\cal Z}_k(\Phi)$ and $U_k(\Phi)$
it is best to choose a non-constant, slowly varying blocked field which
is written as
\eqn\fsumm{\Phi^a(x)=\Phi_0^a+\tphi^a(x),}
with $\Phi_0^a=\Phi_0\delta^{a,1}$.
Simple comparison of \efactt\ and \efact\ gives
\eqn\gle{U_k(\Phi_0)=V(\Phi_0)+{1\over 2\Omega}{\rm Tr'}{\rm ln}K^{ab}
(\Phi_0)}
and
\eqn\wavz{ \tilde {\cal Z}_{a,k}(\Phi_0)=\tilde Z_a(\Phi_0).}
 From now on, quantities with no written arguments are understood to be
evaluated at $\Phi_0$.

In order to obtain higher order correction for the wavefunction
renormalization constant, we incorporate the effect of $\tphi$ up to
quadratic order by writing
\eqn\kexp{\cases{\eqalign{ K^{ab}&=\bigl(K_0^{ab} +\delta K_0^{ab}+\delta
K_1^{ab} +\delta K_2^{ab}\bigr)\delta^{4}(x-y) +O(\tphi^3,\partial^3) \cr
F^{a}&=F_0^{a}+\delta F_1^{a} +\delta F_2^{a}+O(\tphi^3,\partial^3) ,\cr}}}
where
\eqn\knull{\cases{\eqalign{K_0^{ab}&
=-(Z_a\delta^{ab}+Z_a^{(a)}\Phi_0\delta^{a,1}\delta^{b,1})\bx+V^{(ab)} \cr
\delta K_0^{ab}&=\half\Phi_0\Bigl(2Z_a^{(a)}\delta^{a,1}\delta^{b,1}
-\za_b\delta^{b,1}-\zb_a\delta^{a,1}\Bigr)\bx \cr
\delta K_1^{ab}&=-\zc_a\tphi^c\delab\bx-{1\over 2}\Bigl(\za_b\bx\tphi^b
+\zb_a\bx\tphi^a\Bigr)-{1\over 2}\zab_{\ell}\Phi_0\bx\tphi^1 \cr
&~~-{1\over 2}\Bigl\{\za_b\tphi^b+\zb_a\tphi^a+\Phi_0(\delta^{b,1}\zac_b
+\delta^{a,1}\zbc_a)\tphi^c\Bigr\}\bx +V^{(abc)}\tphi^c \cr
\delta K_2^{ab}&= -{1\over 2}\Bigl[\zac_b\tphi^c(\bx\tphi^b
+\tphi^b\bx)+\zbc_a\tphi^c(\bx\tphi^a+\tphi^a\bx)
+\zab_c\tphi^c\bx\tphi^c \cr
&~~+Z^{(cd)}_a\tphi^c\tphi^d\delta^{ab}\bx\Bigr]
+{1\over 2}V^{(abcd)}\tphi^c\tphi^d ,\cr }}}
and
\eqn\fexp{\cases{\eqalign{F_0^a&=-{1\over 2}Z_a\Phi_0\delta^{a,1}\bx
+V^{(a)} \cr
\delta F_1^a &=-{1\over 2}\Bigl[Z_b^{(a)}\Phi_0\delta^{b,1}\bx\tphi^b
+Z_a\bigl(\bx\tphi^a+\tphi^a\bx\bigr)+Z_a^{(b)}\Phi_0\delta^{a,1}\tphi^b\bx
\Bigr]+V^{(ab)}\tphi^b \cr
\delta F_2^{a}&=-{1\over 2}\Bigl[Z^{(a)}_b\tphi^b\bx\tphi^b
+Z_c^{(ab)}\Phi_0\delta^{c,1}\tphi^b\bx\tphi^c+ Z_a^{(b)}\tphi^b
(\bx\tphi^a+\tphi^a\bx) \cr
&\qquad +Z_a^{(bc)}\Phi_0\delta^{a,1}\tphi^b\tphi^c\bx\Bigr]
+{1\over 2}V^{(abc)}\tphi^b\tphi^c .\cr}}}
Note that summation over repeated indices $(a,b,c,d=1,\cdots, N; i,j,k,=
2,\cdots, N)$ is implied, and the
numerical subscripts correspond to the order of $\tphi$.
The effective action can now be written as:
\eqn\effgen{\tilde S_k(\Phi)=S(\Phi_0+\tphi)
+\delta {\tilde S}^1_k(\Phi_0+\tphi)+\delta{\tilde S}^2_k(\Phi_0+\tphi)}
where
\eqn\efs{ S(\Phi_0+\tphi)=\int_x\Bigl(-{Z_a\over2}\tphi^a\bx\tphi^a+V
+V^{(a)}\tphi^a+\half\tphi^a V^{(ab)}\tphi^b+\cdots\Bigr),}
\eqn\efss{\eqalign{\delta{\tilde S}^1_k&=
\half{\rm Tr'}{\rm ln}\Bigl( K_0+\delta K_0+\delta{K_1}+\delta{ K_2}\Bigr)
=\half{\rm Tr'}{\rm ln}\Bigl(K_0+\delta K_0\Bigr) -{1\over 4}{\rm Tr'}\Bigl(
{ K_0}^{-1}\delta {K_1}{K_0}^{-1}\delta {K_1}\Bigr) \cr
&
+\half\sum_{n=0}^{\infty}(-1)^n{\rm Tr'}\Bigl[\Bigl(K_0^{-1}\delta K_0\Bigr)^n
K_0^{-1}\delta K_1\Bigr]
+\half\sum_{n=0}^{\infty}(-1)^n{\rm Tr'}\Bigl[\Bigl(K_0^{-1}\delta K_0\Bigr)^n
K_0^{-1}\delta K_2\Bigr]+\cdots, }}
and
\eqn\esdf{\eqalign{-\delta{\tilde S}_k^2 &={1\over 2}\int_p^{'}
\bigl(F_0+\delta F_1
+\delta F_2\bigr)\bigl(K_0+\delta K_0+\delta K_1+\delta K_2\bigr)^{-1}
\bigl(F_0+\delta F_1+\delta F_2\bigr)+\cdots \cr
&
={1\over 2}\int_p^{'}\Biggl\{ F_0K_0^{-1}\mat F_0+F_0K_0^{-1}\mat\delta F_1\cr
&
+\delta F_1 K_0^{-1}\mat F_0
-F_0K_0^{-2}\delta K_1\matt F_0 \cr
&
+F_0K_0^{-1}\mat\delta F_2+\delta F_2K_0^{-1}\mat F_0 \cr
&
-F_0K_0^{-2}\delta K_1\mat\delta F_1-\delta F_1K_0^{-2}\delta K_1\mat F_0\cr
&
+F_0K_0^{-1}\Bigl[(K_0^{-1}\delta K_1)^2\mattt-(K_0^{-1}K_2)\matt\Bigr]F_0\cr
&
+\delta F_1K_0^{-1}\mat\delta F_1 \Biggr\}+\cdots.}}

To illustrate the above formalism, we consider the $O(N)$
scalar $\lambda\phi^4$ theory defined by
\eqn\lagg{V(\phi)={\mu^2\over 2}\phi^2(x)+{\lambda\over 4!}(\phi^2(x))^2 ,}
where
\eqn\eet{ \phi^2=\phi^a\phi^a. }
With
\eqn\vab{ V^{(ab)}=(\mu^2+{\lambda\over 6}\Phi_0^2)\delta^{ab}
+{\lambda\over 3}\Phi_0^2\delta^{a,1}\delta^{b,1} ,}
the matrix ${ K_0^{ab}}$ in the momentum space representation
has the eigenvalues
\eqn\eigen{{\cal K}^a=\cases{\Delta^{-1}_{\ell}=\tilde\zell p^2+u_{\ell},
&$a=1$\cr
\cr
\Delta^{-1}_t=\tilde Z_tp^2+u_t,&$a=2,\cdots, N,$ \cr}}
\eqn\vv{\cases{\eqalign{u_{\ell}&=\mu^2+{\lambda\over 2}\Phi_0^2 \cr
u_t&=\mu^2 + {\lambda\over 6}\Phi_0^2, \cr }}}
which yields
\eqn\trk{{\rm Tr'}{\rm ln}{ K_0^{ab}}
=\Omega\int_p^{'}\Biggl\{{\rm ln}\Delta^{-1}_{\ell}
+(N-1){\rm ln}\Delta^{-1}_t\Biggr\}.}
Note that \wfrcs\ implies
\eqn\wavee{\cases{\eqalign{\tilde\zell&=\zell+\zell^{(1)}\Phi_0 \cr
\tilde\zell^{(1)}&=2\zell^{(1)}+\zell^{(11)}\Phi_0 \cr
\tilde\zell^{(11)}&=3\zell^{(11)}+\cdots \cr
\tilde\zell^{(i)}&=\zell^{(i)}+\zell^{(1i)}\Phi_0\cr
\tilde\zell^{(ij)}&=\zell^{(ij)}+\cdots,\cr}}}
\eqn\wavet{\cases{\eqalign{\tilde\zt&=\zt \cr
\tilde\zt^{(1)}&=\zt^{(1)}\cr
\tilde\zt^{(i)}&=2\zt^{(i)}. \cr}}}

In evaluating \efss\ and \esdf, we act on the $x$-dependent $\tphi$
in the trace by the derivative operators contained in
$F$ and $K$. The trace is performed in the plane-wave basis
by commuting the momentum operator $p_\mu=i\partial_\mu$
to the right end of the expressions.
The commutation relations utilized for this procedure are tabulated in
Appendix A. Note that whenever an operator $p_\mu$ acts on
$\tphi$ it yields $i\partial_\mu\tphi$ which contributes in
the IR because its Fourier decomposition is
vanishing above the scale $k$. When the operator $p_\mu$
reaches the right end of the expression it gets replaced by the
trace integration variable, $p_\mu$. Although this momentum value is in the
ultraviolet, the contribution represents a simple number which multiplies
the $\tphi$ dependence in the infrared. Therefore, a matrix element
of the form ${\rm Tr'}{\cal F}(\tphi)p_\mu{\cal G}(\tphi)$ can be separated
into ${\rm Tr'}{\cal F}(\tphi)i\partial_\mu({\cal G}(\tphi))$ and
${\rm Tr'}{\cal F}(\tphi){\cal G}(\tphi)p_\mu$ using the commutation
techniques. Iterating this algorithm, the contributions to ${\cal Z}$ which
are of the form ${\rm Tr'}{\cal F}(\tphi)\bx{\cal G}(\tphi)$ can be
isolated. Moreover, since the
blocked action is real we can actually commute the derivative operators
to the left end instead within the trace, as was done in \fraser\ and
adopted here in this paper.

One arrives after lengthy algebra at
\eqn\dkone{\eqalign{\delta K_1^{ab}&=p^2\biggl\{
\Bigl[Z_a^{(c)}\delta^{ab}+{1\over 2}\Phi_0(Z_b^{(ac)}\delta^{b,1}+
Z_a^{(bc)}\delta^{a,1})\Bigr]\tphi^c
+{1\over 2}Z_b^{(a)}\tphi^b+{1\over 2}Z_a^{(b)}\tphi^a\biggr\} \cr
&
-\Bigl[Z_a^{(c)}\delta^{ab}+{1\over 2}\Phi_0(Z_b^{(ac)}\delta^{b,1}+
Z_a^{(bc)}\delta^{a,1})\Bigr]\bx\tphi^c -\Bigl[Z_b^{(a)}\bx\tphi^b
+Z_a^{(b)}\bx\tphi^a\Bigr] \cr
&
-\half\zell^{(ab)}\Phi_0\partial^2\tphi^1
-2i\Bigl[Z_a^{(c)}\delta^{ab}+{1\over 2}\Phi_0(Z_b^{(ac)}\delta^{b,1}+
Z_a^{(bc)}\delta^{a,1})\Bigr]p_{\mu}\parmu\tphi^c \cr
&
+V^{(abc)}\tphi^c-ip_{\mu}\Bigl[Z_b^{(a)}
\parmu\tphi^b+Z_a^{(b)}\parmu\tphi^a\Bigr]+\cdots,}}
\eqn\dwte{\eqalign{\delta K_2^{ab}&={1\over 2} p^2\Bigl\{\zac_b\tphi^b
+\zbc_a\tphi^a+\zcd_a\delta^{ab}\tphi^d\Bigr\}
\tphi^c +{\lambda\over 6}\Bigl(\delab\tphi^c\tphi^c+2\tphi^a\tphi^b\Bigr)\cr
&
-{1\over 2}\bigl[\zac_b\tphi^c\partial^2\tphi^b+\zbc_a\tphi^c\partial^2
\tphi^a+\zab_c\tphi^c\bx\tphi^c\bigr]+\cdots,}}
\eqn\fui{\eqalign{\delta F_1^a&={1\over 2}p^2\Bigl[Z_a\tphia+\zell^{(b)}
\Phi_0\tphib\delta^{a,1}\Bigr]+\delab u_t\tphib+{\lambda\over 3}\Phi_0^2
\tphi^1\delta^{a,1}-Z_a\bx\tphi^a\cr
&
-{1\over 2}\Bigl(Z_b^{(a)}\delta^{b,1}+Z_a^{(b)}\delta^{a,1}
\Bigr)\Phi_0\partial^2\tphib-ip_{\mu}\Bigl[Z_a\parmu\tphia+\zell^{(b)}
\Phi_0\delta^{a,1}\parmu\tphib\Bigr]+\cdots,}}
and
\eqn\fuui{\eqalign{\delta F_2^a &= {1\over 2}p^2\bigl[Z_a^{(b)}\tphia
+\zell^{(bc)}\Phi_0\tphic\delta^{a,1}\bigr]\tphib-{1\over 2}\bigl[Z_a^{(b)}
\tphi^a+Z_b^{(a)}\tphib+\zell^{(ab)}\Phi_0\tphi^1\bigr]\bx\tphib \cr
&
+{1\over 2}V^{(abc)}\tphib\tphic +\cdots.}}
As for the calculation of $\delta{\tilde S}^2_k$,
since no $x$ integration occurs in
\esdf, it is only the Fourier transform of $\tphi$. If $\tphi(p)$ is
constrained such that $p < k$, then the contribution of
$\delta{\tilde S}^2_k$ to the blocked action vanishes since
the $p$ integration is performed over the range $k > p$.

In order to identify the contribution to the wavefunction renormalization
constants and the blocked potential we now rewrite the blocked action
$\tilde S_k(\Phi)$ as:
\eqn\blocc{ \tilde S_k(\Phi_0+\tphi)=\int_x\Biggl( -{1\over 2}{\cal Z}_k^{ab}
\tphi^a\bx\tphi^b + U_k+(U_k^{(a)},\tphi^a)
+{1\over 2}(\tphi^a,~U_k^{(ab)}\tphi^b)+\cdots\Biggr),}
from which one obtains the $\tilde\phi$-independent blocked potential:
\eqn\effz{\eqalign{U_k&=
V + {1\over 2}{\rm Tr'}{\rm ln}\Bigl(K_0+\delta K_0\Bigr) \cr
&
=V+{1\over 2}\int_p^{'}\Bigl({\rm ln}\Delta_{\ell}^{-1}+(N-1){\rm ln}
\Delta_t^{-1}\Bigr).}}
For the wavefunction renormalization constants, we first
collect terms proportional to $\tphi^1\bx\tphi^1$ for computing
${\cal Z}_{k,\ell}$ in the longitudinal direction. Use of
\wavee\ and \wavet\ yields (see Appendix B):
\eqn\lon{\eqalign{{\tilde {\cal Z}}_{k,\ell}&=\tilde\zell+
{1\over 2}\int_p^{'}
\Bigl\{a_{11}+a'_{11}+\tilde a_{11}+a^{*}_{11}-2b_{11}\Bigr\} \cr
&
=\tilde\zell+{1\over 2}\int_p^{'}\Biggl\{
\dell^2\Bigl[\tilde\zell\uell\dell^2\lambda^2\Phi_0^2-\zell^{(1)}\lambda
\Phi_0(1-\uell\dell+2\uell^2\dell^2)-{(\zell^{(1)})}^2p^2(1+\uell^2\dell^2)
\Bigr] \cr
&
+(N-1)\ut\delt^3\Bigl\{\delt\Bigl[\tilde Z_t{\lambda^2\over 9}\Phi_0^2
-(Z_t^{(1)})^2p^2\ut\Bigr]+{\lambda\over 3}\Phi_0 Z_t^{(1)}\bigl(1-2\ut
\delt\bigr)\Bigr\} \cr
&
-\zell^{(ii)}\Phi_0\delt^2\bigl(Z_t^{(1)}p^2+{\lambda\over 3}\Phi_0\bigr)
-{1\over 4}{({\tilde\zell^{(i)}})}^2p^2\dell\delt\Bigl(4+\uell^2\dell^2+
\ut^2\delt^2\Bigr) \cr
&
+{\theta\over 4}\zell^{(i)}
\Phi_0\dell\delt p^2\Bigl[\zell^{(j)}\zell^{(ji)}\Phi_0\delt p^2-8
\zell^{(i1)}\Bigr]+3\zell^{(11)}\theta\dell+\zell^{(ii)}\delt \Biggr\}.}}
One may replace the $Z$ factors above by the corresponding $\tilde Z'$s
since the difference
involves contributions having more than two orders of derivative in the $Z's$.
Finally, to obtain $\tilde{\cal Z}_{k,\ell}(\Phi)$, we simply replace
$\Phi_0$ in $\tilde{\cal Z}_{k,\ell}(\Phi_0)$ by
$\Phi$, with implied ``normal ordering'' such that all $p$ dependences
be moved to the front of the $\Phi$-dependent expressions.
As noted in \fraser, there is no ambiguity in this procedure provided
that we carefully compare the terms in the expansions of \efactt\ and
\effgen.

In a similar manner, we can write down the $\Phi$-dependent
wavefunction renormalization
constant in the transverse ${ij}$ direction as:
\eqn\tran{\eqalign{{\tilde {\cal Z}}_{k,t(ij)}
&=\tzt+{1\over 2}\int_p^{'}
\Bigl\{a_{ij}+a'_{ij}+\tilde a_{ij}+a^{*}_{ij}-2b_{ij}\Bigr\} \cr
&
=\tzt +{1\over 2}\int_p^{'}\Biggl\{ -{\tilde\zell^{(i)}}{\tilde\zell^{(j)}}
p^2\Bigl[\uell^2\dell^4+\delt^2\bigl(3+{2N+3\over 2}\ut^2\delt^2\bigr)\Bigr]\cr
&
-(\tilde Z_t^{(k)})^2\delta^{ij}p^2\delt^2\bigl(2+u_t^2\delt^2\bigr)
+{1\over 6}\tilde\zell^{(ij)}\Phi_0\dell\delt\Bigl\{-3\tilde Z_t^{(1)}
p^2\bigl(2+\uell^2\dell^2+\ut^2\delt^2\bigr) \cr
&\qquad
+\lambda\Phi\Bigl[\uell\dell
(1-2\uell\dell)+\ut\delt(1-2\ut\delt)\Bigr]\Bigr\}\cr
&
+{\lambda\over 3}\Phi\delta^{ij}\dell\delt\Bigl[{\lambda\over 3}
\Phi\bigl(\tilde\zell
\uell\dell^2+\tzt\ut\delt^2\bigr)-{\tzt^{(1)}\over 2}\bigl(4-\uell\dell
-\ut\delt+2\uell^2\dell^2+2\ut^2\delt^2\bigr)\Bigr] \cr
&
-{p^2\over 4}\dell\delt\Bigl[{({\tilde Z_t^{(1)}})}^2\delta^{ij}\bigl(
4+\uell^2\dell^2+\ut^2\delt^2\bigr)+\tilde\zell^{(ik)}\tilde\zell^{(kj)}\Phi^2
\bigl(\uell^2\dell^2+\ut^2\delt^2\bigr)\Bigr] \cr
&
+\delta^{ij}\Bigl[\theta\dell \tilde Z_t^{(11)}+\tilde Z_t^{(kk)}\delt
+{\theta\over 4}\tilde \zell^{(k)}\Phi p^2\dell\delt\Bigl(\tilde
\zell^{(\ell)}\tilde Z_t^{({\ell}k)}\Phi p^2\delt-4\tilde Z_t^{(k1)}\Bigr)
\Bigr]\cr
&
+2\delt\Bigl[\tilde Z_t^{(ij)}+{\theta\over 2}\tilde\zell^{(i)}\Phi p^2\dell
\Bigl({1\over 2}\tilde \zell^{(k)}\tilde Z_t^{(kj)}\Phi p^2\delt-
\tilde Z_t^{(j1)}\Bigr)\Bigr]\Biggr\}.}}
Consider the limiting case in which the derivative terms of the
$\tilde Z's$ can be neglected, the above
expressions can be reduced to:
\eqn\locs{U_k=V +\half\int_p^{'}\Biggl\{{\rm ln}\Bigl(1+{\uell\over p^2}\Bigr)
+(N-1){\rm ln}\Bigl(1+{u_t\over p^2}\Bigr) \Biggr\} ,}
\eqn\ders{\eqalign{\tilde{\cal Z}_{k,\ell}&=\tilde\zell+{\lambda^2\Phi^2
\over 2}\int_p^{'}\biggl\{\tilde\zell\uell\dell^4+{N-1\over 9}\tilde Z_t u_t
\delt^4\biggr\} \cr
&
=\tilde\zell+{\lambda^2\Phi^2\over 192\pi^2}\Biggl[{{(3\tilde\zell k^2
+\uell)\uell}\over {\tilde\zell(\tilde\zell k^2+\uell)^3}}+{N-1\over 9}
{{(3\tilde Z_tk^2+u_t)u_t}\over{\tilde Z_t(\tilde Z_tk^2+ u_t)^3}}\Biggr] ,}}
and
\eqn\derr{ \tilde {\cal Z}_{k,t} =\tilde Z_t ,}
in agreement with that obtained in \fraser\ for the one-component case.
\medskip
\bigskip
\centerline{\bf IV. RENORMALIZATION GROUP FLOW EQUATIONS}
\medskip
\nobreak
\xdef\secsym{4.}\global\meqno = 1
\medskip
Equation \effz\ gives the contribution of the modes between
$k<p<\Lambda$ to the blocked action in the one-loop independent
mode approximation since the systematic feedbacks from the high modes to the
low ones are neglected. In order to improve upon such approximation,
we first consider the
case when the cutoff is changed infinitesimally from $k\to k-\Delta k$,
leading to:
\eqn\loggr{ k\partial_k U_k=-{k^4\over16\pi^2}\Biggl\{{\rm ln}
\Bigl[1+{{\lambda\Phi^2/2}\over {\tilde\zell k^2+\mu^2}}\Bigr]+(N-1)
{\rm ln}\Bigl[1+{{\lambda\Phi^2/6}\over {\tilde Z_t k^2+\mu^2}}\Bigr]
\Biggr\},}
which is a linear partial differential equation.
The equation is not yet suitable for the systematical repetition of the
elimination of the modes since the right-hand side of \loggr\ is derived by
using the specific potential, \lagg. Since elimination of modes
changes the specific structure of the lagrangian, it is
better to start the whole computation
with a general potential. Upon replacing the $\Phi$-dependent terms
on the right hand side of \loggr\ by
\eqn\frr{\cases{\eqalign{ {\upp}(\Phi)&={{\partial^2 U_k(\Phi)}\over
{\partial\Phi_1^2}}\cr
{\utt}(\Phi)&={{\partial^2 U_k(\Phi)}\over
{\partial\Phi_2^2}},\cr}}}
one obtains a new RG equation:
\eqn\lorr{k\partial_k U_k(\Phi)=-{k^4\over16\pi^2}\Biggl\{{\rm ln}
\Bigl[{{\tilde\calzl(\Phi)k^2+{\upp}(\Phi)}\over
{\tilde\calzl(0)k^2+{\upp}(0)}}\Bigr]+(N-1)
{\rm ln}\Bigl[{{\tilde\calzt(\Phi)k^2+{\utt}(\Phi)}\over
{\tilde\calzt(0)k^2+{\utt}(0)}}\Bigr]\Biggr\},}
which accumulates the effects of the eliminated modes
in a systematic way as we lower the cutoff. Contrary to \loggr,
eq. \lorr\ is now a nonlinear partial differential flow equation.
In the same manner, we can write down the corresponding RG
flow equations for the $\tilde{\cal Z}'$s:
\eqn\lonk{\eqalign{k\partial_k{\tilde{\cal Z}}_{k,\ell} =&
-{k^4\over 16\pi^2}\Biggl\{
\dellk^2\Bigl[\tilde\zellk\uellk\dellk^2({U^{(111)}_k})^2
-\tilde\zellk^{(1)}U^{(111)}_k\bigl(1-\uellk\dellk+2(\uellk)^2\dellk^2\bigr)\cr
&
-{(\tilde\zellk^{(1)})}^2k^2\bigl(1+(\uellk)^2\dellk^2\bigr)
\Bigr]+(N-1)U_k^{(22)}\deltk^3\Bigl\{ \tilde Z_{k,t}^{(1)}
U_k^{(221)}\bigl(1-2\utk\deltk\bigr) \cr
&
+\deltk\Bigl[\tilde Z_{k,t}\bigl(U_k^{(221)}\bigr)^2-(\tilde Z_{k,t}^{(1)})^2
U_k^{(22)}k^2\Bigr] \Bigr\}-\tilde\zellk^{(ii)}\Phi\deltk^2\bigl(\tilde
Z_{k,t}^{(1)}k^2 +U_k^{(221)}\bigr)\cr
&
-{k^2\over 4}{({\tilde\zellk^{(i)}})}^2\dellk\deltk\Bigl(4+(\uellk)^2\dellk^2+
(\utk)^2\deltk^2\Bigr)+\tilde\zellk^{(ii)}\deltk \cr
&
+{\theta_k\over 4}\tilde\zellk^{(i)}
\Phi\dellk\deltk k^2\Bigl[\tilde\zellk^{(j)}\tilde\zellk^{(ji)}\Phi\deltk k^2
-8\tilde\zellk^{(i1)}\Bigr]+3\tilde\zellk^{(11)}\theta_k\dellk \Biggr\},}}
and
\eqn\trank{\eqalign{k\partial_k{\tilde{\cal Z}}_{k,t(ij)}&=
-{k^4\over 16\pi^2}\Biggl\{ -{\tilde\zellk^{(i)}}{\tilde\zellk^{(j)}}k^2
\Bigl[(\uellk)^2\dellk^4+\deltk^2\bigl(3+{2N+3\over 2}(\utk)^2\deltk^2\bigr)
\Bigr]\cr
&\qquad\qquad
-(\tilde Z_{k,t}^{(k)})^2\delta^{ij}k^2\deltk^2\bigl(2+(\utk)^2\deltk^2
\bigr) \cr
&
+{1\over 6}\tilde\zellk^{(ij)}\Phi\dellk\deltk\Bigl\{-3\tilde Z_{k,t}^{(1)}
k^2\bigl(2+(\uellk)^2\dellk^2+(\utk)^2\deltk^2\bigr) \cr
&
+U_k^{(111)}\Bigl[\uellk\dellk
\bigl(1-2\uellk\dellk\bigr)+\utk\deltk\bigl(1-2\utk\deltk\bigr)\Bigr]\Bigr\}\cr
&
+U_k^{(221)}\delta^{ij}\dell\delt\Bigl[U_k^{(221)}\bigl(\tilde\zellk
\uellk\dellk^2+\tz_{k,t}\utk\deltk^2\bigr) \cr
&
-{\tz_{k,t}^{(1)}\over 2}
\bigl(4-\uellk\dellk-\utk\deltk+2(\uellk)^2\dellk^2+2(\utk)^2\deltk^2\bigr)
\Bigr] \cr
&
-{k^2\over 4}\dellk\deltk\Bigl[{({\tilde Z_{k,t}^{(1)}})}^2\delta^{ij}\bigl(
4+(\uellk)^2\dellk^2+(\utk)^2\deltk^2\bigr) \cr
&\qquad\qquad
+\tilde\zellk^{(ik)}\tilde
\zellk^{(kj)}\Phi^2\bigl((\uellk)^2\dellk^2+(\utk)^2\deltk^2\bigr)\Bigr] \cr
&
+\delta^{ij}\Bigl[\theta_k\dellk \tilde Z_{k,t}^{(11)}+\tilde
Z_{k,t}^{(kk)}\deltk
+{\theta_k\over 4}\tilde\zellk^{(k)}\Phi k^2\dellk\deltk(\tilde
\zellk^{(\ell)}\tilde Z_{k,t}^{({\ell}k)}\Phi k^2\deltk-4\tilde
Z_{k,t}^{(k1)})\Bigr]\cr
&
+2\deltk\Bigl[\tilde Z_{k,t}^{(ij)}+{\theta_k\over 2}\tilde\zellk^{(i)}\Phi
k^2\dellk
\Bigl({1\over 2}\tilde \zellk^{(k)}\tilde Z_{k,t}^{(kj)}\Phi k^2\deltk-
\tilde Z_{k,t}^{(j1)}\Bigr)\Bigr]\Biggr\},}}
where
\eqn\eigg{ \theta_k= \Bigl[1-{1\over 4}(k^2)^2\dellk\deltk(\tilde
\zellk^{(i)}\Phi)^2\Bigr]^{-1} ,}
and
\eqn\propag{\cases{\eqalign{ \Delta_{k,\ell}^{-1} &=\tilde\calzl(\Phi)
k^2+\upp(\Phi) \cr
\Delta_{k,t}^{-1}&=\tilde\calzt(\Phi)k^2+\utt(\Phi) . \cr}}}

So far the only approximation we made was to make truncation in the derivative
expansion up to $O(\partial^2)$. The complicated flow equations can
further be simplified
if one also truncates in the amplitude of the fluctuations which is
another expansion parameter of the Landau-Ginsburg method.
By taking a constant background $\Phi_0$ along the longitudinal direction
and replacing $\Phi_0$ by the general inhomogeneous $\Phi$, we have
\eqn\ukk{\cases{\eqalign{ U_k^{(11)}&
=2(\udot+2\uddot\Phi^2) \cr
U_k^{(22)}&=2\udot \cr
U_k^{(111)}&=4(3\uddot+2\udddot\Phi^2)\vec\Phi \cr
U_k^{(221)}&=4\uddot\vec\Phi, \cr}}}
and
\eqn\gfd{\cases{\eqalign{ {\tilde{\cal Z}}_k^{(1)}&
=2{{\tilde{\cal Z}}}^{'}_k
\vec\Phi \cr
{\tilde{\cal Z}}_k^{(i)}&=0 \cr
{\tilde{\calz}}_k^{(11)}&=4({{\tilde{\cal Z}}}^{''}_k\Phi^2
+{{\tilde{\cal Z}}}^{'}_k) \cr
{\tilde{\cal Z}}_k^{(ij)}&=2{{\tilde{\cal Z}}}^{'}_k\delta^{ij}, \cr}}}
where $\vec\Phi$ points in the $1$ direction and the prime denotes
differentiation with respect to $\Phi^2$.
%
%
The three coupled non-linear differential RG flow equations now become
(see Appendix B)
\eqn\lou{\eqalign{k\partial_k U_k(\Phi)
&=-{k^4\over16\pi^2}\Biggl\{{\rm ln}
\Bigl[{{\tilde\calzl(\Phi)k^2+2\bigl(U'_k(\Phi)+2U''_k(\Phi)\Phi^2\bigr)}\over
{\tilde\calzl(0)k^2+2U'_k(0) } }\Bigr]\cr
&+(N-1){\rm ln}\Bigl[{{\tilde\calzt(\Phi)k^2+2U'_k(\Phi)}\over
{\tilde\calzt(0)k^2+2U'_k(0)}}\Bigr]\Biggr\},}}
\eqn\lonkt{\eqalign{k\partial_k{\tilde{\cal Z}}_{k,\ell} &=
-{k^4\over 16\pi^2}\Bigl( a_{k,\ell}+a'_{k,\ell}+\tilde a_{k,\ell}
+ a^{*}_{k,\ell}-2b_{k,\ell}\Bigr) \cr
&
=-{k^4\over 16\pi^2}\Biggl\{ 4\dellk^2\Phi^2 \Biggl(-{(\tc'_{k,\ell})}^2
k^2\Bigl[1+4\dellk^2\bigl(U_k'+2U_k''\Phi^2\bigr)^2\Bigr] \cr
&
+8\tc_{k,\ell}\dellk^2\bigl(U_k'+2U_k''\Phi^2\bigr)\bigl(3U''_k+2U_k'''
\Phi^2\bigr)^2 \cr
&
-2\tc'_{k,\ell}\bigl(3U_k''+2U_k'''\Phi^2\bigr)\Bigl[1-2(U_k'+2U_k''\Phi^2)
\dellk+8(U_k'+2U_k''\Phi^2)^2\dellk^2\Bigr]\Biggr) \cr
&
+4(N-1)\deltk^2\Phi^2\biggl(4U_k'\deltk\bigl[\deltk\bigl(
2\tilde{\cal Z}_{k,t}(U_k'')^2-
(\tilde{\cal Z}_{k,t}')^2k^2U_k'\bigr)+\tilde{\cal Z}_{k,t}'U_k''
(1-4U_k'\deltk)\bigr] \cr
&
-\tc'_{k,\ell}\bigl(\tc'_{k,t}k^2+2U_k''\bigr)\biggr)
+12\bigl(\tc''_{k,\ell}\Phi^2+\tc'_{k,\ell}\bigr)\dta_{k,\ell}
+2\bigl(N-1\bigr)\tc'_{k,\ell}\dta_{k,t} \Biggr\},}}
and
\eqn\trankt{\eqalign{k\partial_k{\tilde{\cal Z}}_{k,t}&=
-{k^4\over 16\pi^2}\Bigl( a_{k,t}+a'_{k,t}+\tilde a_{k,t}+a^{*}_{k,t}
-2b_{k,t}\Bigr)\cr
&
=-{k^4\over 16\pi^2}\Biggl\{ 2\dellk\deltk\Phi^2\Biggl(
-\tilde{\cal Z}'_{k,\ell}k^2\Bigl[\tc'_{k,t}+2(U_k')^2\deltk^2\bigl(
\tc'_{k,\ell}+2\tc'_{k,t}\bigr)\Bigr] \cr
&
+16\tc_{k,t}U_k'(U_k'')^2\deltk^2
-(\tc'_{k,t})^2k^2\Bigl(1+2(U_k')^2\deltk^2\Bigr)
+4\tc'_{k,\ell}\deltk U_k'U_k''\bigl(1-4U_k'\deltk\bigr) \cr
&
-4\tc'_{k,t}U_k''\Bigl[1-U_k'\deltk+4(U_k')^2\deltk^2\Bigr]
+4U_k''\biggl[
4\tc_{k,\ell}U_k''\bigl(U_k'+2U_k''\Phi^2\bigr)\dellk^2 \cr
&
-\tc'_{k,t}\Bigl(1-(U_k'+2U_k''\Phi^2)\dellk+4(U_k'+2U_k''\Phi^2)^2\dellk^2
\Bigr)\biggr] \cr
&
+4\tc'_{k,\ell}U_k''\dellk\bigl(U_k'+2U_k''\Phi^2\bigr)\Bigl[1-4(U_k'
+2U_k''\Phi^2)\dellk\Bigr] \cr
&
-\tc'_{k,\ell}k^2\biggl[\tc'_{k,t}+2\bigl(\tc'_{k,\ell}+2\tc'_{k,t}\bigr)
\bigl(U_k'+2U''_k\Phi^2\bigr)^2\dellk^2\biggr] \Biggr)\cr
&
+4\bigl(\tc''_{k,t}\Phi^2
+\tc'_{k,t}\bigr)\dta_{k,\ell}+2\bigl(N+1\bigr)\tc'_{k,t}\dta_{k,t}\Biggr\}
.}}
In the limiting case where $N=1$ and the derivative couplings are
neglected, we have
\eqn\loone{ k\partial_k{\tilde{\cal Z}}_{k,\ell} =
-{2k^4\over\pi^2}{\tilde{\cal Z}}_{k,\ell}\Phi^2\Delta_{k,\ell}^4
\bigl(U'_k+2U''_k\Phi^2\bigr)\bigl(3U''_k+2U'''_k\Phi^2\bigr)^2.}
\medskip
\bigskip
\centerline{\bf V. ASYMPTOTIC SCALINGS}
\medskip
\nobreak
\xdef\secsym{5.}\global\meqno = 1
\medskip
\noindent{\bf a. Ultraviolet regime}
\medskip
As can be seen from \ukk, the $O(N)$ scalar model possesses a natural
mass scale:
\eqn\masssc{m_R^2={\partial^2\over\partial\Phi^2_1}U_{k=0}(\sigma)
=4\sigma^2U''_k,}
where $\sigma=<\Phi>$, the vacuum expectation value.
This mass scale is what separates the UV
from the IR regimes. For sufficiently deeply in the UV
or the IR regime, and in the linearizable vicinity of the
fixed point(s) one finds asymptotic scaling. The calculation presented
above reproduces the expected perturbative results in the UV scaling
regime. To make contact with the usual RG equation where only the
relevant and the marginal operators associated with the
UV fixed point are followed, we turn to the RG
coefficient functions for the longitudinal modes:
\eqn\betaff{\beta_n(k)=k\partial_kg_n,}
where
\eqn\ccdim{g_n=\partial^n_{\Phi_1}U_k(\sigma).}
In order to find the true critical
exponents we have to make the coupling constants dimensionless by the
help of $k$. To this end we introduce
\eqn\betafdl{\tilde\beta_n(k)=k\partial_k\tilde g_n,}
with
\eqn\ccdiml{\tilde g_n=k^{n-4}\partial^n_{\Phi_1} U_k(\sigma).}
The naive power counting for determining the sign of the critical exponents
is especially transparent in this case. In fact, \betaff\ now becomes
\eqn\betafdlf{\tilde\beta_n(k)=k^{n-4}(k\partial_k+n-4)g_n,}
where $k\partial_kg_n$ is treated as small in perturbation expansion.
Thus, $g_n$ is classified as being relevant or irrelevant for $n<4$ or $n>4$,
respectively.

In the case of spontaneous symmetry breaking for which $\sigma\ne 0$,
the evolution equations of the quadratic and quartic coupling
constants take on the forms
\eqn\ftwo{\eqalign{\beta_2&=\gamma_m =k\partial_k
\bigl(4\sigma^2U''_k(\sigma)\bigr) \cr
&
=-{k^4\over 4\pi^2}
\Biggl\{\dellk\bigl(3U''_k+12\sigma^2U'''_k+4\sigma^2U''''_k\bigr)
-3\sigma^2\dellk^2\bigl(3U''_k+2\sigma^2U'''_k)^2 \cr
&
+(N-1)\Bigl[\deltk\bigl(U''_k+2\sigma^2U'''_k\bigr)
-4\sigma^2\deltk^2(U''_k)^2\Bigr]\Biggr\},}}
and
\eqn\ffou{\eqalign{\beta_4&=k\partial_k\Bigl[4
\bigl(3U''_k+12\sigma^2U'''_k+4\sigma^4U''''_k\bigr)\Bigr] \cr
&
=-{3k^4\over 2\pi^2}\Biggl\{ 5\dellk U'''_k -2\dellk^2\Bigl(9(U''_k)^2
+192\sigma^2U''_kU'''_k+224\sigma^4(U'''_k)^2\Bigr)\cr
&
+96\dellk^3\sigma^2\bigl(U''_k+4\sigma^2U'''_k\bigr)\bigl(3U''_k
+2\sigma^2U'''\bigr)^2
+64\dellk^4\sigma^4\bigl(3U''_k+2\sigma^2U'''_k\bigr)^4 \cr
&
+(N-1)\Bigl[ \deltk U_k'''
-2\deltk^2\Bigl((U''_k)^2+12\sigma^2U_k''U_k'''+4\sigma^4(U_k''')^2\Bigr)\cr
&
+32\deltk^3\sigma^2(U_k'')^2\bigl(U''_k+2\sigma^2U'''_k\bigr)
-64\deltk^4\sigma^4(U_k'')^4 \Bigr]+\cdots\Biggr\},}}
where the ellipsis denotes $O(U^{(4)}_k)$ terms, and the inverse propagators
now become:
\eqn\propg{\cases{\eqalign{ \Delta_{k,\ell}^{-1} &=\tilde\calzl
k^2+m_R^2 \cr
\Delta_{k,t}^{-1}&=\tilde\calzt k^2 . \cr}}}
In addition, there are also flow equations for terms which are odd in
$\sigma$ due to spontaneous breaking of symmetry. For example, we have
\eqn\fthr{\eqalign{\beta_3&=k\partial_k\Bigl[4\sigma
\bigl(3U''_k+2\sigma^2U'''_k\bigr)\Bigr] \cr
&
=-{{\sigma k^4}\over 2\pi^2}\Biggl\{ 15\dellk U'''_k -
18\dellk^2(3U_k''+2\sigma^2U_k''')(U_k''+4\sigma^2U_k''')
+16\dellk^3\sigma^2(3U''_k+2\sigma^2U'''_k)^2 \cr
&
+(N-1)\deltk\Bigl[ 3 U_k'''-6\deltk U_k''(U''_k+2\sigma^2U_k''')
+16\deltk^2\sigma^2(U_k'')^3 \Bigr]+\cdots\Biggr\}.}}

The RG coefficient functions for the wave function renormalization constants
are given as:
\eqn\wfcf{\eqalign{\gamma_{k,\ell}&={\tilde{\cal Z}}^{-1}_{k,\ell}
{}~k\partial_k{\tilde{\cal Z}}_{k,\ell}\Big\vert_{\sigma} \cr
&
=-{{\tilde {\cal Z}_{k,\ell}}^{-1}k^4\over 16\pi^2}\Biggl\{
4\dellk^2\sigma^2 \Biggl(-{(\tc'_{k,\ell})}^2
k^2\Bigl[1+16(U_k'')^2\dellk^2{(\sigma^2)}^2\Bigr] \cr
&
-2\tc'_{k,\ell}\bigl(3U_k''+2U_k'''\sigma^2\bigr)\Bigl[1-4U_k''\dellk\sigma^2
+32(U_k'')^2\dellk^2{(\sigma^2)}^2\Bigr] \cr
&
+16\tc_{k,\ell}U_k''\dellk^2\sigma^2\bigl(3U''_k+2U_k'''
\sigma^2\bigr)^2\Biggr)
-4(N-1)\tc'_{k,\ell}(\tilde{\cal Z}_{k,t}'k^2+2U_k'')\deltk^2\sigma^2 \cr
&
+12\bigl(\tc''_{k,\ell}\sigma^2+\tc'_{k,\ell}\bigr)\dta_{k,\ell}
+2\bigl(N-1\bigr)\tc'_{k,\ell}\dta_{k,t} \Biggr\},}}
and
\eqn\wfcfu{\eqalign{\gamma_{k,t}&={\tilde{\cal Z}}^{-1}_{k,t}
{}~k\partial_k{\tilde{\cal Z}}_{k,t} \cr
&
=-{{{\tilde {\cal Z}_{k,t}}^{-1}~k^4}\over 16\pi^2}\Biggl\{2\dellk\deltk
\sigma^2\Biggl(-\tc'_{k,t}k^2\bigl(2\tc'_{k,\ell}+\tc'_{k,t}\bigr) \cr
&
+8U_k''\Bigl\{4\tc_{k,\ell}(U_k'')^2\dellk^2\sigma^2-\tc'_{k,t}\bigl[
1-U''_k\dellk\sigma^2+8(U_k'')^2\dellk^2{(\sigma^2)}^2\bigr]\Bigr\} \cr
&
+8\tc'_{k,\ell}(U_k'')^2\dellk\sigma^2\Bigl[1-8U_k''\dellk\sigma^2
-(\tc'_{k,\ell}+2\tc'_{k,t})k^2\dellk\sigma^2\Bigr] \Biggr) \cr
&
+4\bigl(\tc''_{k,t}\sigma^2
+\tc'_{k,t}\bigr)\dta_{k,\ell}+2\bigl(N+1\bigr)\tc'_{k,t}\dta_{k,t}\Biggr\}
.}}

Using $U_k''={\lambda_R/12}$ by neglecting
$O(U'''_k)$, i.e., dropping the irrelevant coupling constants in the
UV scaling regime, and setting $m_R^2=\lambda_R\sigma^2/3$,
the above expressions become:
\eqn\ftwos{\beta_2=-{\lambda_Rk^4\over 16\pi^2}\Biggl\{\Bigl(\dellk
+{N-1\over 3}\deltk\Bigr)-\lambda_R\sigma^2\Bigl(\dellk^2+{N-1\over 9}
\deltk^2\Bigr)\Biggr\},}
\eqn\fthrs{ \beta_3={{\sigma\lambda_R^2 k^4}\over 48\pi^2}
\Biggl\{3\dellk^2\Bigl(3-2\lambda_R\sigma^2\dellk\Bigr)+(N-1)\deltk^2
\Bigl(1-{2\lambda_R\over 9}\sigma^2\deltk\Bigr)\Biggr\},}
\eqn\ffous{\eqalign{\beta_4&={\lambda_R^2k^4\over 16\pi^2}\Biggl\{3
\dellk^2\Bigl[1-4\lambda_R\sigma^2\dellk+2\lambda_R^2\sigma^4\dellk^2\Bigr]\cr
&
+{(N-1)\over 27}\deltk^2\Bigl[1-12\lambda_R\sigma^2\deltk
+2\lambda_R^2\sigma^4\deltk^2\Bigl]\Biggr\},}}
\eqn\wfcfs{\eqalign{\gamma_{k,\ell}&={\tilde{\cal Z}}^{-1}_{k,\ell}
{}~k\partial_k{\tilde{\cal Z}}_{k,\ell} \cr
&
=-{{{\tilde {\cal Z}_{k,\ell}}^{-1}~k^4}\over 16\pi^2}\Biggl\{
4\dellk^2\sigma^2 \Biggl(-{(\tc'_{k,\ell})}^2
k^2\Bigl[1+{\lambda^2\over 9}\dellk^2{(\sigma^2)}^2\Bigr]
+{\lambda^3\over 12}\tc_{k,\ell}\dellk^2\sigma^2\cr
&
-{\lambda\over 2}\tc'_{k,\ell}\Bigl[1-{\lambda\over 3}\dellk\sigma^2
+{2\lambda^2\over 9}\dellk^2{(\sigma^2)}^2\Bigr]\Biggr)
-4(N-1)\tc'_{k,\ell}(\tilde{\cal Z}_{k,t}'k^2+{\lambda\over 6})
\deltk^2\sigma^2 \cr
&
+12\bigl(\tc''_{k,\ell}\sigma^2+\tc'_{k,\ell}\bigr)\dta_{k,\ell}
+2\bigl(N-1\bigr)\tc'_{k,\ell}\dta_{k,t} \Biggr\},}}
\eqn\wfcfu{\eqalign{\gamma_{k,t}&={\tilde{\cal Z}}^{-1}_{k,t}
{}~k\partial_k{\tilde{\cal Z}}_{k,t} \cr
&
=-{{{\tilde {\cal Z}_{k,t}}^{-1}~k^4}\over 16\pi^2}\Biggl\{ 2\dellk\deltk
\sigma^2\Biggl(-\tc'_{k,t}k^2\bigl(2\tc'_{k,\ell}+\tc'_{k,t}\bigr) \cr
&
+{2\lambda\over 3}\Bigl[{\lambda^2\over 36}\tc_{k,\ell}\dellk^2\sigma^2
-\tc'_{k,t}\bigl(1-{\lambda\over 12}\dellk\sigma^2+{\lambda^2\over 18}
\dellk^2{(\sigma^2)}^2\bigr)\Bigr] \cr
&
+{\lambda^2\over 18}\tc'_{k,\ell}\dellk\sigma^2\Bigl[1-{2\lambda\over 3}
\dellk\sigma^2-(\tc'_{k,\ell}+2\tc'_{k,t})k^2\dellk\sigma^2\Bigr] \Biggr) \cr
&
+4\bigl(\tc''_{k,t}\sigma^2
+\tc'_{k,t}\bigr)\dta_{k,\ell}+2\bigl(N+1\bigr)\tc'_{k,t}\dta_{k,t}\Biggr\}
.}}

The leading order part of the above coefficient functions
for $k^2/m_R^2\to\infty$ and $\tilde\calzl,\tilde\calzt=1+O(\lambda_R)$:
\eqn\prot{\cases{\eqalign{ \beta_2 &= -{N+2\over 48\pi^2}
\lambda_Rk^2\cr
\beta_3 &={\sigma\lambda_R^2\over 48\pi^2}\bigl(N+8\bigr)\cr
\beta_4 &={\lambda_R^2\over 16\pi^2}\bigl({N+8\over 27}\bigr)\cr
\gamma_{k,\ell}&=-{\lambda_R^3\over 48\pi^2}\bigl({\sigma^2\over k^2}\bigr)^2
\longrightarrow 0 \cr
\gamma_{k,t} &=-{\lambda_R^3\over 48\pi^2}\bigl({\sigma^2\over 3k^2}\bigr)^2
\longrightarrow 0. \cr}}}
The last two lines are in agreement with the well known fact that there
is no need for wave function renormalization at the one-loop order. In fact,
their coefficient functions are vanishing in the ultraviolet. When
symmetry breaking appears,  $\tilde\calzl$ and $\tilde\calzt$
evolve differently from each other as we leave the UV regime.

If one wishes to compare the above results with the usual $\epsilon$
expansion in critical $\lambda\phi^4$ theory in $4-\epsilon$ dimension,
it is sufficient to consider a symmetrical theory in which one
sets $U'_k(0)=0$ and defines a dimensionless
coupling constant $\tilde\lambda_k=12k^{-\epsilon}U''_k(0)
=12k^{-\epsilon}\lambda_k$. From \ffou, one is led to
\eqn\ffour{\beta_4 =k\partial_k {\tilde\lambda_k}
=-\epsilon\tilde\lambda_k+{3\tilde\lambda_k^2\over 16\pi^2}
\Bigl[{1\over {\tilde\calzl}}+{N-1\over 9}{1\over {\tilde\calzt}}\Bigr]
-{3U'''_k(0)k^2\over 16\pi^2}\Bigl[{5\over {\tilde\calzl}}
+{N-1\over {\tilde\calzt}}\Bigr],}
which for $\tilde\calzl=\tilde\calzt=1$ reproduces the leading order
results obtained by standard perturbation method.
\medskip
\medskip
\noindent{\bf b. Infrared regime}
\medskip
The scaling is more involved in the IR regime where the expansion
$k^2<<m^2_R$ should be applied \lp. The strong non-linearities
of the RG equation prevented us from constructing
the corresponding scaling operators. Instead we only argue that
relevant operators must exist in the IR scaling regime.

Let us first begin with the naive argument outlined in the Introduction
where $\tilde {\cal Z}$ is set to be unity.
In fact, due to Goldstone's theorem which asserts
$U^{(22)}_{k=0}(\sigma)=U'_{k=0}(\sigma)=0$, we take
\eqn\aszp{U^{(22)}_k(\sigma)=O(k^\rho)}
with $\rho=2$. In this limit with $U^{(22)}_k(\sigma)=ck^2$, the leading
order contribution to the $\beta$ function becomes
\eqn\betafi{\eqalign{\tilde\beta_n(k)&={(-1)^nk^n\over 16\pi^2}
\biggl({U^{(122)}_k(\sigma)\over k^2+U^{(22)}_k(\sigma)}
\biggr)^n(1+O(k^2))+(n-4)\tilde g_n\cr
&\approx{(-1)^n\over16\pi^2}\biggl({U^{(122)}_k(\sigma)\over(1+c)k}
\biggr)^n(1+O(k^2)),}}
where
\eqn\trlder{U^{(122)}_k(\sigma)={\partial^3U_k(\sigma)\over
\partial\Phi^2_2\partial\Phi_1}=4\sigma U_k''={\lambda_R\over 3}\sigma+\cdots.}
Since $U^{(122)}_k(\sigma)$ is finite for non-critical system inside the
symmetry broken phase the $\beta$ functions develop power-like IR divergences
when $U^{(122)}_k(\sigma)>0$.

This simple argument relies on the leading order of the derivative
expansion. Having gone through the computation of the wave function
renormalization constants we can verify that our conclusion remains
valid in the next order of the derivative expansion, too. In particular,
we show that the IR power singularities of the $\beta$ functions
persist for the RG equation \lou\-\trankt.
In the presence of a nontrivial wave function renormalization constant
the $\beta$ function reads as
\eqn\betafiz{\eqalign{\tilde\beta_n(k)&={(-1)^nk^n\over 16\pi^2}
\biggl({\tilde\calzt^{(1)}(\sigma)k^2+U^{(122)}_k(\sigma)\over\tilde
\calzt(\sigma)k^2+U^{(22)}_k(\sigma)}\biggr)^n(1+O(k^2))+(n-4)\tilde g_n\cr
&\approx{(-1)^n\over 16\pi^2}
\biggl({U^{(122)}_k(\sigma)\over k^{1-\nu}}\biggr)^n(1+O(k^2)),}}
where the asymptotic behavior
\eqn\aszt{\tilde\calzt=O(k^{-\nu})}
$\nu \ge 0$ was assumed.

In order to show that $\nu<1$, which would imply the persistence of
IR singularity in \betafiz, we verify the consistency of \lonkt\ and
\trankt\ by assuming \aszp, \aszt\ and
\eqn\asl{\tilde\calzl=O(k^{-\mu}).}
A similar power-law dependence on $k$ also applies to the
$\tilde{\cal Z}'_k$'s and the $\tilde{\cal Z}''_k$'s since differentiation
with respect to $\Phi^2$ does not affect the power counting in $k$.
While $\dellk$ remains finite as $k\to 0$ due to the presence of
mass gap, the transverse propagator is scaled as
\eqn\pros{\Delta_{k,t}=O(k^{\nu-2}).}
Substituting the above scalings in \lonkt\ and \trankt\
for matching the leading singularities gives
\eqn\muineq{ -\mu={\rm min}\Bigl\{~6-2\mu, ~4-\mu,~-4+3\nu+\rho,
2-\mu+\nu, ~-2+2\nu+\rho, \cdots\Bigr\},}
%
%
%
and
\eqn\nuineq{\eqalign{-\nu={\rm min}\Bigl\{& ~4-\mu,~4-\nu, ~-2\mu+3\nu+2\rho,
{}~-2+2\nu+\rho,~-\mu+2\nu+\rho,~ 2, \cr
&
-2-\mu+3\nu+2\rho,~\nu+\rho,~2+\rho,~2-\mu+\nu,~4-2\mu+\nu+\cdots\Bigr\}.}}
%
%

Relying on the method of independent-mode approximation in \lp, we have
$\rho=2$ which implies $\mu=2$ and $\nu=0$. Thus, we conclude that presence of
IR singularity in the $\beta$ function \betafiz\ remains unchanged after taking
into consideration the wavefunction renormalization constants.
Note that the longitudinal wavefunction renormalization constant
$\calz_{\ell}$ is found to be quadratically divergent in the IR
limit. This is in contradiction with the usual assumption \ref\itzykson\
\eqn\spectras{0< {\cal Z}< 1 ,}
made in Minkowski space-time. Thus the invariance of $\calz$ under the Wick
rotation, the one-loop RG equation, i.e. $\zeta\approx0$ and
\spectras\ are inconsistent. It is not clear how to describe the vacuum but
a deviation from the weakly coupled perturbative scenario is expected due to
the Goldstone modes.

We emphasize again that the difference between our $\beta$ functions
\ftwo, \ffou\ and the usual ones in \prot\ stem from the
terms $O(\sigma^2/k^2)$ which are neglected at the UV
fixed point. These are just the pieces which make the UV
and the IR scaling laws different.

In the course of investigating the IR scaling behavior, our computations
are free of IR divergences. This is because for
a finite value of the adjustable cutoff, $k$, the
elimination of the degrees of freedom in \efact\ contains the integration
for modes with finite momentum $k<p<\Lambda$ and the usual singularities
of the massless theories do not show up. They appear only when $k\to0$,
the limit where singularity appears in the $\beta$ functions. The
determination of IR scaling operators is rather involved since
it requires a complete resummation of the singularities which
emerge in that limit.
\medskip
\bigskip
\centerline{\bf VI. SUMMARY}
\medskip
\nobreak
\xdef\secsym{6.}\global\meqno = 1
\medskip

Most of the applications of the RG are related to
the investigation of the impact of local field operators
which are introduced into the theory in the UV regime.
The concept of universality which is being supported by the
linearized RG equations
can be used as long as the phenomena we are
interested lie within the linearizability of the UV
fixed point. Such phenomena can therefore be described by a simple
hamiltonian containing few relevant operators.
This certainly is the case for the critical, i.e. massless
$\phi^4$ model near four dimensions where both the UV and the
IR fixed points are Gaussian. The classification of the
scaling operators is valid for all length scales in this model.

The situation is radically different for massive theories. There we have
two energy scales, the UV cutoff and the mass gap. There is
no reason to expect the same scaling laws at both sides of the
mass gap. In fact, the
renormalized trajectory runs towards large values of the mass squared
as we approach the IR regime. As the vacuum expectation value
of the field reaches $O(1/\epsilon)$, where $\epsilon=4-d$ then
the three-point vertex becomes of order $1$ in the symmetry broken phase
and the expansion around the
Gaussian fixed point is not applicable.

This is the generic situation
for high-energy physics where the renormalized trajectory passes by the
vicinities of different UV fixed points and we find different
scaling laws at different energy ranges. The usual concept of
universality is not applicable here since the relevant coupling
constants of an UV fixed point parametrize only the physics
of a given energy range. The trajectory is driven away from the region
of the linearizability by the relevant or
the irrelevant coupling constants as we move towards lower or
higher energies, respectively. For the sake of definiteness we considered
the $O(N)$ model in this paper which has a single finite scale and
exhibits only two fixed points, an UV and an IR.

It is usually claimed that there is only one (completely trivial)
relevant coupling
constant at the IR fixed point, the mass. But this claim ignores
the IR divergences of the massless theories which
may generate relevant operators as the observational scale approaches the
IR regime. Another class of models where the IR scaling might
be rather nontrivial is where a symmetry is broken spontaneously.
This case is interesting because it emphasizes the importance of
\eqn\rat{\zeta={2\pi\over Lk} ={\kappa\over L},}
the ratio of the observational length scale, $\kappa=2\pi k^{-1}$, and the IR
cutoff, $L$. For $L$ close to the characteristic mass scale of the theory
there is no symmetry breaking and the evolution
equation for the effective coupling constants reflects the symmetrical
dynamics. For large but finite $L$ the symmetry is still preserved
but the spontaneous symmetry breaking scheme becomes a good approximation
for the dynamics. Spontaneous breakdown of symmetry can take place
only asymptotically in the limit $L\to\infty$. Since $\kappa$
can never exceed $L$ for a finite system, we have $\zeta<1$.
Therefore, pattern of symmetry breaking
can be uncovered in the evolution equation only for $\zeta\sim0$. In fact,
for $\zeta\sim1$, one detects the symmetry-restoring long-range slow
fluctuations, and the observables at this energy scale truly reflect the
symmetrical dynamics. Thus the characteristic size of the system $\sim L$
should be much larger than the observational length scale $\kappa$
in order to recover the usual picture of symmetry breaking.

There is another rather technical reason for staying in the region of small
$\zeta$. The
higher loop contributions to the RG equations are suppressed
by the inverse of the number of the modes in the blocked system which
is $O(\zeta^d)$. Thus the studies of systems undergoing
spontaneous symmetry breaking using the one-loop RG equation requires the
removal of the IR cutoff by sending $L\to\infty$ before $\kappa\to\infty$,
thereby making $\zeta=0$.

The locality is lost at the IR fixed point which may lead to the breakdown
of the derivative expansion and global scaling operators. All we know is
that classical physics is recovered
at the IR fixed point of massless theories. In this case
the soft particle emission allows the spread of the energy from
the microscopical to macroscopical length scales. On the other hand,
when the theory possesses
a mass gap, the energy cannot be distributed to arbitrary
long distances and the IR physics is still controlled
by coherent quantum effects, e.g., superconductivity.

We found IR singularities in the one-loop $\beta$ function for the odd
vertices of the $O(N)$ model when the first two orders
in the derivative expansion of the renormalized lagrangian were retained.
This supports the notion of strong coupling IR physics of the Goldstone
modes. The amplification of the effective coupling strengths
can be understood by recalling that the ``restoring force'' for the
fluctuations, i.e., the eigenvalue of the small fluctuation operator
is vanishing in the IR limit of a massless theory.
Consequently large fluctuations are always present in the IR regime
and invalidate the expansion methods.

The limitation of the concept of universality due to the
existence of several fixed points can be nicely demonstrated in the
$O(N)$ model. Consider the coupling constant of an odd vertex in the
spontaneously broken phase. Being irrelevant in the vicinity of the
UV fixed point, it decreases as we move in the
IR direction. Its value should be small when we reach the
crossover region between the UV and IR scaling, at
the mass gap. Universality, i.e., the insensitivity on the
irrelevant initial
conditions of the renormalized trajectory seems to be holding down to this
energy scale. But as we continue our journey in the IR direction
our coupling constant starts to grow. Although we could ignore it on the UV
side of the mass gap, it plays an important role on the IR end.
Furthermore, its actual value may depend strongly on the
ultraviolet initial value of the renormalized trajectory. The suppression
which produced the universal behavior down to the mass gap turns out to be
an amplification in the IR scaling regime
and the UV value of this coupling constant
influences the long-distance features of the model. This possibility raises
questions on the sufficiency of renormalized field theories in
describing low energy phenomena.

We believe that only few nonrenormalizable operators become important
in this manner. To demonstrate this point consider the Standard Model.
Despite all complications in the IR regime the experiments
performed in the vicinity of the crossover, $O(100GeV)$,
can be parametrized by the help of the renormalizable coupling constants.
Where then is the room for the possible violation of universality ?
The conjectured strong coupling physics in the IR plays an important
part in forming the vacuum but remains virtually invisible at higher
energy. The only important parameters they provide are the values
of the condensates. These condensates appear under the disguise of
renormalizable coupling constants, say lepton masses in the
usual scheme. But the relation
between the mass and the condensate bridges the energy scale of the
mass and zero and thus its tree-level form is highly questionable.
On the one hand, the order parameters of the spontaneously broken
symmetries are formed in the asymptotical IR regime. On the other
hand, they parametrize the effective vertices at the crossover.

It is reasonable
to expect that the only impact of the IR modes on the
physics of the higher energy processes is the generation of the
symmetry breaking condensates. Thus universality actually holds when the
physics is parametrized by the help of the renormalizable coupling
constants and the condensates. It is useless if we make an
attempt to derive the values of the condensates starting with
the UV parameters, from one fixed point only.

This scenario leaves universality unharmed
for ferromagnets. In fact, the condensate of the nonlinear $\sigma$-
model is a unit vector and there is no possibility of
changing its length. In contrast of this situation, the physics of the
superconductors may show non-universal features. In particular,
the supercurrent density might depend on non-renormalizable
coupling constants of QED which are provided by theories of
higher energy scale, such as the Standard Model. In turn, the Higgs
condensate of the Standard Model is a non-universal function of the
bare parameters of a GUT, etc.

One would object the speculations about relevant operators
for the IR fixed point of a superconductor since there
is no gap in the physical spectrum. But the massless excitations
are present in the gauge-dependent sector where the Higgs mechanism
relegates them and continues to influence the dynamics of the gauge invariant
modes. Their presence can be seen from the long range
confining forces acting between two magnetic charges. In fact, in the
absence of massless modes all interactions are screened. The
situation is similar to the QCD vacuum where the long-range
confining modes coexist with finite range the Yukawa forces due to the
massive glueball exchanges.

In closing we repeat again that our results rely on the derivative expansion.
It would be of key importance to support or disclaim its validity
for the four-dimensional models in the IR regime.
\medskip
\centerline{\bf ACKNOWLEDGEMENTS}
\medskip
\nobreak
J. P. thanks J. Zinn-Justin and V. Branchina for stimulating discussions.
\medskip
\bigskip
\centerline{\bf APPENDIX A}
\medskip
\nobreak
\xdef\secsym{{\rm A}.}\global\meqno = 1
\medskip
We collect here the commutation relations which were used in deducing
the wavefunction renormalization constants. All the relations derived in
this Appendix are based upon the simple rule
\eqn\comz{ [~\tphi,~p_{\mu}~]~= -i\parmu \tphi .}
Note that the appearance of the negative sign which differs from the
conventional definition is due to the fact that the space-time traces are
evaluated in the plane-wave basis with all $p$ dependences being moved
to the {\it left} of $x$-dependent field operators. Repeated use of the
above gives
\eqn\com{ [~\tphi,~p^2~]~= -\bx \tphi- 2ip_{\mu}\parmu \tphi }
\eqn\coms{ [~\tphi,~\pmu p^2~]~= -\bigl(\pmu\bx\tphi+2\pnu\parnu\parmu\tphi
\bigr)-i\bigl(p^2\parmu\tphi+2\pmu\pnu\parnu\tphi\bigr)+\cdots}
\eqn\coms{ [~\tphi,~{(p^2)}^2~]~= -3p^2\bx \tphi- 4ip^2p_{\mu}\parmu \tphi
+\cdots}
\eqn\comm{ [~[~\tphi,~p^2~],~p^2~]~=
- 4p_{\mu}p_{\nu}\partial_{\mu}\partial_{\nu}\tphi +\cdots}
\eqn\commu{\eqalign{ [~\tphi, ~\Delta~]&= \tilde Z\Delta^2[~p^2,~\tphi~]
+\tilde Z^2\Delta^3 [~p^2,~[~p^2, ~\tphi~]~]+ \cdots \cr
&
= \tilde Zu\Delta^3\bx \tphi+2i\tilde Z\Delta^2p_{\mu}\parmu \tphi
+\cdots }}
\eqn\commu{ [~\tphi^a\tphi^b,~p^2~]=[~\tphi^a\tphi^b, ~\Delta~]=0}
\eqn\commut{ [~\tphi, ~p^2\Delta~]= -\tilde Z^{-1}u [~\tphi,~\Delta~]
=-u^2\Delta^3\bx \tphi-2iu\Delta^2 p_{\mu}\parmu \tphi+\cdots}
\eqn\commuti{ [~\tphi, ~\Delta p_{\mu}~] =
\tilde Z\Delta^2\bigl(u\Delta p_{\mu}\bx \tphi+2p_{\nu}\parnu\parmu\tphi
\bigr)+i\Delta\bigl(2Z\Delta p_{\mu}p_{\nu}\parnu\tphi-\parmu\tphi\bigr)
+\cdots}
\eqn\commuta{\eqalign{ [~\tphi,~{(p^2)}^2\Delta~]&= \tilde Z^{-1}[~\tphi,
{}~p^2~]+\tilde Z^{-2}u^2[~\tphi,~\Delta~] \cr
&
=-p^2\Delta\bigl(1+u\Delta+u^2\Delta^2\bigr)\bx \tphi-2ip^2\Delta
(1+u\Delta)p_{\mu}\parmu \tphi+\cdots}}
\eqn\commuto{[~\tphi,~p^2\Delta p_{\mu}~] =-u\Delta^2\bigl(u\Delta p_{\mu}
\bx \tphi+2p_{\nu}\parnu\parmu\tphi\bigr)-i\Delta\bigl(2u\Delta
p_{\mu}p_{\nu}\parnu\tphi+p^2\parmu\tphi\bigr)+\cdots.}
\eqn\commuta{ [~\tphi,~\Delta_{\ell}\delt~]
=\dell\delt\Bigl\{ \bigl({\tilde Z}_tu_t\delt^2+{\tilde Z}_{\ell}u_{\ell}
\dell^2-{\tilde Z}_{\ell}{\tilde Z}_t\dell\delt p^2\bigr)\bx\tphi
+2i\bigl({\tilde Z}_{\ell}\dell+{\tilde Z}_t\delt\bigr)
p_{\mu}\parmu\tphi \Bigr\}+\cdots}
\eqn\couta{ [~\tphi,~\dta^2~]=-\tz\dta^3\bigl(1-3u\dta\bigr)\bx\tphi
+4i\tz\dta^3\pmu\parmu\tphi+\cdots}
\eqn\commuta{\eqalign{ [~\tphi,~p^2\Delta_{\ell}\delt~]&=\dell\delt\Bigl\{
\bigl({\tilde Z}_tp^2u_t\delt^2+{\tilde Z}_{\ell}p^2u_{\ell}
\dell^2-u_{\ell}u_t\dell\delt\bigr)\bx\tphi \cr
&\qquad\qquad
+2i\bigl(1-u_{\ell}\dell-u_t\delt\bigr)p_{\mu}\parmu\tphi\Bigr\}+\cdots}}
\eqn\commuta{ [~\tphi,~p^2\dta^2~]=u\dta^3\bigl(2-3u\dta\bigr)\bx\tphi
+2i\dta^2\bigl(1-2u\dta\bigr)\pmu\parmu\tphi+\cdots}

After generating the derivative terms with the above commutation
relations, one may untangle the $x-$ and $p$-dependent terms with the
following useful relations:
\eqn\sdrte{\eqalign{\bigl(p^2 f_1-2ip_{\mu}\parmu f_1
-\bx f_2+ f_3\bigr)\dta&=\dta\Bigl\{p^2 f_1-\bx f_2+f_3 +\tz\dta\bigl[
p^2(1+u\dta\bigr)\bx f_1+u\dta\bx f_3\bigr] \cr
&\quad
+2i\pmu\dta\bigl[-u\parmu f_1+\tz\parmu f_3\bigr]\Bigr\}+\cdots ,}}
\eqn\sdrte{\eqalign{ &\bigl(p^2 f_1-2ip_{\mu}\parmu f_1
-\bx f_2+ f_3\bigr)\Delta\bigl(p^2 g_1-2i\pnu\parnu g_1
-\bx g_2+ g_3\bigr)\cr
&
=\dta\biggl\{ p^2\bigl(p^2f_1+f_3\bigr)g_1+\bigl(p^2f_1+f_3\bigr)g_3
-\bigl(p^2f_1+f_3\bigr)\bx g_2-\bigl(p^2g_1+g_3\bigr)\bx f_2 \cr
&
+\bigl[p^2\bigl(2-u^2\dta^2\bigr)g_1+\tz p^2\dta(1+u\dta)g_3\bigr]\bx f_1
+\bigl[(1+u\dta-u^2\dta^2)g_1+\tz u\dta^2g_3\bigr]\bx f_3 \cr
&
+2i\pmu\dta\bigl(p^2g_1+g_3\bigr)\bigl(-u\parmu f_1+\tz\parmu f_3\bigr)
+\cdots\biggr\},}}
which for $g=f$ reduces to
\eqn\sdrg{\eqalign{ &\bigl(p^2 f_1-2ip_{\mu}\parmu f_1
-\bx f_2+ f_3\bigr)\Delta\bigl(p^2 f_1-2i\pnu\parnu f_1
-\bx f_2+ f_3\bigr)\cr
&
=\dta\biggl\{ p^2\bigl(p^2f_1+f_3\bigr)f_1+\bigl(p^2f_1+f_3\bigr)f_3
-2\bigl(p^2f_1+f_3\bigr)\bx f_2 \cr
&
+p^2\bigl(2-u^2\dta^2\bigr)f_1\bx f_1+\tz\dta^2u f_3\bx f_3
+(2+u\dta-2u^2\dta^2)f_3\bx f_1 \cr
&
+2i\pmu\dta\bigl(p^2f_1+f_3\bigr)\bigl(-u\parmu f_1+\tz\parmu f_3\bigr)
+\cdots\biggr\}.}}
Similarly, we have:
\eqn\sdrtee{\eqalign{ &\bigl(p^2 f_1-2ip_{\mu}\parmu f_1
-\bx f_2+ f_3\bigr)\bigl(p^2 g_1-2i\pnu\parnu g_1
-\bx g_2+ g_3\bigr)={(p^2)}^2f_1 g_1+p^2(f_1g_3+f_3g_1) \cr
&
+f_3g_3-2i\pmu(p^2g_1+g_3)\parmu f_1 +(p^2f_1+f_3)\bigl(\bx g_1-\bx g_2\bigr)
-(p^2g_1+g_3)\bx f_2+\cdots,}}
and
\eqn\frte{\eqalign{ \bigl(&f_1-\bx f_2-2i\pmu\parmu f_3\bigr)\bigl[g_0
+{(p^2)}^2g_1+p^2g_2-\bx g_3-2ip^2\pnu\parnu g_4-2i\pnu\parnu g_5\bigr] \cr
&
=f_1g_0+{(p^2)}^2f_1g_1+p^2f_1g_2-\bx f_2 g_0-p^2\bigl(3\bx f_1+p^2\bx f_2
+2p^2\bx f_3\bigr) \cr
&
-\bigl(\bx f_1+p^2\bx f_2+p^2\bx f_3\bigr)g_2-f_1\bx g_3
+p^2\bigl(3f_1+p^2f_3\bigr)\bx g_4+(2f_1+p^2f_3\bigr)\bx g_5+\cdots.}}

\medskip
\bigskip
\centerline{\bf APPENDIX B}
\medskip
\nobreak
\xdef\secsym{{\rm B}.}\global\meqno = 1
\medskip

In computing the effective blocked action $\tilde S_k$, one encounters
the $N\times N$ matrix $M$ of the form:
\eqn\mons{M=1+\km\delta K_0=\left(\matrix{1& a^T\cr b& I\cr}\right),}
where
\eqn\mor{ a^T=\left(\matrix{a_{\ell}\zell^{(2)}&
\ldots&a_{\ell}\zell^{(N)}\cr}\right)\qquad a_{\ell}\equiv {1\over 2}p^2
\Phi_0\dell,}
and
\eqn\vve{ b=\pmatrix{a_t\zell^{(2)}\cr\vdots\cr a_t\zell^{(N)}\cr}
\qquad~\qquad a_t\equiv {1\over 2}p^2\Phi_0\delt .}
One can easily verify that its inverse $M^{-1}$ takes on the form:
\eqn\invs{M^{-1}=\left(\matrix{\theta& -\theta a^T\cr -\theta b&
I+\theta ba^T\cr}\right),}
where
\eqn\eigg{ \theta={\bigl(1-a^Tb\bigr)}^{-1}=\bigl[1-a_{\ell}a_t
{(\zell^{(i)})}^2\bigr]^{-1}.}
Employing the relations:
\eqn\trex{ {\rm Tr'}{\rm ln} \Bigl(1+{ K_0}^{-1}\delta K \Bigr)
= \sum_{n=1}^{\infty}{(-1)^{n+1}\over n}{\rm Tr'}\Bigl\{
\bigl({ K_0}^{-1}\delta { K}\bigr)^n \Bigr\} }
and
\eqn\fosu{\sum_{n=0}^{\infty}(-1)^n{\rm Tr'}\Bigl[\Bigl(K_0^{-1}\delta
K_0\Bigr)^n K_0^{-1}\delta K_{\alpha}\Bigr]={\rm Tr'}\Bigl[\Bigl(1
+K_0^{-1}\delta K_0\Bigr)^{-1}
K_0^{-1}\delta K_{\alpha}\Bigr],}
we readily obtain
\eqn\efss{\eqalign{\delta{\tilde S}^1_k&=
\half{\rm Tr'}{\rm ln}\Bigl( K_0+\delta K_0+\delta{K_1}+\delta{ K_2}\Bigr)
=\half{\rm Tr'}{\rm ln}\Bigl(K_0+\delta K_0\Bigr) -{1\over 4}{\rm Tr'}\Bigl(
{ K_0}^{-1}\delta {K_1}{K_0}^{-1}\delta {K_1}\Bigr) \cr
&
+\half\sum_{n=0}^{\infty}(-1)^n{\rm Tr'}\Bigl[\Bigl(K_0^{-1}\delta K_0\Bigr)^n
K_0^{-1}\delta K_1\Bigr]
+\half\sum_{n=0}^{\infty}(-1)^n{\rm Tr'}\Bigl[\Bigl(K_0^{-1}\delta K_0\Bigr)^n
K_0^{-1}\delta K_2\Bigr]+\cdots \cr
&
=\half{\rm Tr'}{\rm ln}\Bigl(K_0+\delta K_0\Bigr) -{1\over 4}{\rm Tr'}\Bigl(
{ K_0}^{-1}\delta {K_1}{K_0}^{-1}\delta {K_1}\Bigr)
+\half {\rm Tr'}\Bigl[\Bigl(1+K_0^{-1}\delta K_0\Bigr)^{-1}
K_0^{-1}\delta K_1\Bigr] \cr
&
+\half {\rm Tr'}\Bigl[\Bigl(1+K_0^{-1}\delta K_0\Bigr)^{-1}
K_0^{-1}\delta K_2\Bigr]+\cdots,}}
In terms of matrix elements,
\eqn\trtr{\eqalign{ {\rm Tr'}\bigl[\bigl(1+K_0^{-1}\delta K_0\bigr)^{-1}
K_0^{-1}\delta K_{\alpha}\bigr] &=\int_x\int_p^{'}\Biggl\{\theta
\dell(\delta K_{\alpha})^{11}+\theta\delt a_{\ell}a_t\zell^{(i)}
\zell^{(j)}(\delta K_{\alpha})^{ji}\Bigr] \cr
&\qquad\qquad
+\delt(\delta K_{\alpha})^{ii}-\theta\bigl(a_{\ell}\delt+a_t\dell\bigr)
\zell^{(i)}(\delta K_{\alpha})^{i1}\Biggr\} ,}}
and
\eqn\treel{\eqalign{{\rm Tr'}\Bigl({ K_0}^{-1}\delta { K_1}{K_0}^{-1}
\delta {K_1}\Bigr)&= \int_x\int_p^{'}\Biggl\{\dell (\delta K_1)^{11}\dell
(\delta K_1)^{11}+\delt(\delta K_1)^{ij}
\delt(\delta K_1)^{ji}\cr
&
+\Bigl[\dell (\delta K_1)^{1i}\delt (\delta K_1)^{i1}
+ \delt (\delta K_1)^{i1}\dell (\delta K_1)^{1i}\Bigr]\Biggr\}.}}
The complicated commutator algebra can be simplified by noting
that the matrix elements take on the forms:
\eqn\enntr{\cases{\eqalign{{(\delta K_1)}^{ab}&=p^2{\bf {\alpha_1}}
-2ip_{\mu}\parmu{\bf {\alpha_1}}-\bx{\bf {\alpha_2}}+{\bf {\alpha_3}} \cr
{(\delta K_2)}^{ab}&=p^2{\bf {\beta_1}}\tphi^c-{\bf {\beta_2}}\bx\tphic
+{\bf {\beta_3}}, \cr}}}
\eqn\coeef{{(\delta K_1)}^{11}:{\cases{\eqalign{\alpha_1&
=\tilde\zell^{(1)}\tphi^1+\tilde\zell^{(i)}\tphi^i \cr
\alpha_2&={3\over 2}\tilde\zell^{(1)}\tphi^1+\tilde\zell^{(i)}\tphi^i \cr
\alpha_3&=\lambda\Phi_0\tphi^1 \cr}}}\qquad
{(\delta K_1)}^{1i}:{\cases{\eqalign{
\tilde\alpha_1&={1\over 2}\bigl(\tilde\zell^{(i)}\tphi^1
+\tilde Z_t^{(1)}\tphi^i+\tilde\zell^{(ij)}\Phi_0\tphi^j\bigr) \cr
\tilde\alpha_2&=\tilde\zell^{(i)}\tphi^1
+\tilde Z_t^{(1)}\tphi^i+{1\over 2}\tilde\zell^{(ij)}\Phi_0\tphi^j \cr
\tilde\alpha_3&={\lambda\over 3}\Phi_0\tphi^i \cr}}}}
\eqn\csds{{(\delta K_1)}^{ij}:\cases{\eqalign{\alpha'_1&=Z_t^{(c)}\delta^{ij}
\tphi^c+ {1\over 2}\bigl(Z_t^{(i)}\tphi^j+Z_t^{(j)}\tphi^i\bigr) \cr
\alpha'_2&=Z_t^{(c)}\delta^{ij}\tphi^c
+ Z_t^{(i)}\tphi^j+Z_t^{(j)}\tphi^i+{1\over 2}\zell^{(ij)}\Phi_0\tphi^1 \cr
\alpha'_3&={\lambda\over 3}\Phi_0\delta^{ij}\tphi^1 .\cr}}}
\eqn\coeef{{(\delta K_2)}^{11}:{\cases{\eqalign{\beta_1&=
\zell^{(1c)}\tphi^1+{1\over 2}\zell^{(cd)}\tphi^d \cr
\beta_2&=\zell^{(1c)}\tphi^1+{1\over 2}Z_c^{(11)}\tphic \cr
\beta_3&={\lambda\over 6}\bigl(\tphic\tphic+2\tphi^1\tphi^1\bigr) \cr}}}}
\eqn\cofr{{(\delta K_2)}^{1i}:{\cases{\eqalign{\tilde\beta_1&=
{1\over 2}\bigl(Z_t^{(1c)}\tphi^i+\zell^{(ic)}\tphi^1\bigr) \cr
\tilde\beta_2&={1\over 2}\bigl(Z_t^{(1c)}\tphi^i+\zell^{(ic)}\tphi^1
+Z_c^{(1i)}\tphic\bigr) \cr
\tilde\beta_3&={\lambda\over 3}\tphi^1\tphi^i \cr}}}}
\eqn\cbtf{{(\delta K_2)}^{ij}:\cases{\eqalign{\beta'_1&={1\over 2}\bigl(
Z_t^{(ic)}\tphi^j+Z_t^{(jc)}\tphi^i+Z_t^{(cd)}\delta^{ij}\tphi^d\bigr) \cr
\beta'_2&={1\over 2}\bigl(Z_t^{(ic)}\tphi^j+Z_t^{(jc)}\tphi^i
+Z_c^{(ij)}\tphic\bigr) \cr
\beta'_3&={\lambda\over 6}\bigl(\delta^{ij}\tphic\tphic+2\tphi^i\tphi^j
\bigr).\cr}}}
Substituting \coeef\ - \cbtf\ into \trtr\ for $\delta K_2$ yields:
\eqn\yyu{ {\rm Tr'}\bigl[\bigl(1+K_0^{-1}\delta K_0\bigr)^{-1}
K_0^{-1}\delta K_2\bigr] =\int_x\int_p^{'}\Bigl\{b_{11}\tphi^1\bx
\tphi^1+b_{ij}\tphi^i\bx\tphi^j \Bigr\}+\cdots,}
where
\eqn\regt{\cases{\eqalign{b_{11}&=-{1\over 2}\bigl(3\zell^{(11)}\theta
\dell+\zell^{(ii)}\delt\bigr)-{\theta\over 8}\zell^{(i)}\Phi_0\dell\delt
p^2\bigl[\zell^{(j)}\zell^{(ji)}\Phi_0\delt p^2-8\zell^{(i1)}\bigr] \cr
b_{ij}&=-{1\over 2}\delta^{ij}\Bigl[\theta\dell Z_t^{(11)}+Z_t^{(kk)}
\delt+{\theta\over 4}\zell^{(k)}\Phi_0p^2\dell\delt\bigl(
\zell^{(\ell)}Z_t^{({\ell}k)}\Phi_0p^2\delt-4Z_t^{(k1)}\bigr)\Bigr]\cr
&\qquad -\delt\Bigl[Z_t^{(ij)}+{\theta\over 2}\zell^{(i)}\Phi_0p^2\dell
\bigl({1\over 2}\zell^{(k)}Z_t^{(kj)}\Phi_0p^2\delt-Z_t^{(j1)}
\bigr)\Bigr] .\cr}}}

For the $\delta K_1$-dependent terms, after much
tedious algebra with the help of the relations found in Appendix A, we have:
\eqn\daon{\eqalign{\dell(\delta K_1)^{11}\dell(\delta & K_1)^{11} =
\dell^2\Bigl\{ {(p^2)}^2\alpha_1^2+2p^2\alpha_1\alpha_3+\alpha_3^2
+2p^2\alpha_1(\bx\alpha_1-\bx\alpha_2)-2\alpha_3\bx\alpha_2 \cr
&
-p^2\uell^2\dell^2\alpha_1\bx\alpha_1+\tz\uell\dell^2\alpha_3\bx\alpha_3
+(2+\uell\dell-2\uell^2\dell^2)\alpha_3\bx\alpha_1 +\cdots\Bigr\} \cr
&
=a_{11}\tphi^1\bx\tphi^1+a_{ij}\tphi^i\bx\tphi^j+\cdots ,}}
\eqn\cof{\cases{\eqalign{a_{11}&=-\dell^2\biggl\{ {(\zell^{(1)})}^2p^2(1+
\uell^2\dell^2)+\zell^{(1)}\lambda\Phi_0(1-\uell\dell+2\uell^2\dell^2)
-\tilde\zell\uell\dell^2\lambda^2\Phi_0^2\biggr\} \cr
a_{ij}&=-{\zell^{(i)}}{\zell^{(j)}} p^2\uell^2\dell^4 ,\cr}}}
\eqn\dao{\delt(\delta K_1)^{ij}\delt(\delta K_1)^{ji}
=a'_{11}\tphi^1\bx\tphi^1+a'_{ij}\tphi^i\bx\tphi^j+\cdots ,}
\eqn\cof{\cases{\eqalign{a'_{11}&=(N-1)\ut\delt^3\Bigl\{\delt\Bigl[
{\lambda^2\over 9}\Phi_0^2\tz_t-(Z_t^{(1)})^2p^2\ut\Bigr]
+{\lambda\over 3}\Phi_0 Z_t^{(1)}(1-2\ut\delt)\Bigr\} \cr
&\qquad-\zell^{(ii)}\Phi_0\delt^2\bigl(Z_t^{(1)}p^2+{\lambda\over 3}\Phi_0
\bigr) \cr
a'_{ij}&=-Z_t^{(i)}Z_t^{(j)}p^2\delt^2\Bigl(3+{2N+3\over 2}u_t^2
\delt^2\Bigr)-{1\over 2}(Z_t^{(k)})^2\delta^{ij}p^2\delt^2\bigl(2+u_t^2\delt^2
\bigr) ,\cr}}}
\eqn\daonn{\dell(\delta K_1)^{1i}\delt(\delta K_1)^{i1}=
\tilde a_{11}\tphi^1\bx\tphi^1+\tilde a_{ij}\tphi^i\bx\tphi^j,}
\eqn\cof{\cases{\eqalign{\tilde a_{11}&=
-{1\over 4}{({\tilde\zell^{(i)}})}^2p^2\dell\delt\bigl(2+\ut^2\delt^2\bigr)\cr
\tilde a_{ij}&=\dell\delt\Biggl\{
{1\over 6}\tilde\zell^{(ij)}\Phi_0\Bigl[-3\tilde Z_t^{(1)}p^2(1+
\ut^2\delt^2)+\lambda\Phi_0\ut\delt(1-2\ut\delt)\Bigr] \cr
&
-\delta^{ij}\Bigl[{1\over 4}{({\tilde Z_t^{(1)}})}^2p^2
(2+\ut^2\delt^2)-{\lambda\over 3}\Phi_0\Bigl({\lambda\over 3}\Phi_0
\tz_t\ut\delt^2-{\tz^{(1)}_t\over 2}(2-\ut\delt+2\ut^2\delt^2)\Bigr)\Bigr]\cr
&
-{1\over 4}\tilde\zell^{(ik)}\tilde\zell^{(kj)}\Phi_0^2p^2\ut^2\delt^2
\Biggr\} ,\cr}}}
\eqn\daonn{\delt(\delta K_1)^{i1}\dell(\delta K_1)^{1i}=
a^{*}_{11}\tphi^1\bx\tphi^1+a^{*}_{ij}\tphi^i
\bx\tphi^j,}
\eqn\cof{\cases{\eqalign{ a^{*}_{11}&=-{1\over 4}{({\tilde\zell^{(i)}})}^2
p^2\dell\delt\bigl(2+\uell^2\dell^2\bigr)\cr
a^{*}_{ij}&=\dell\delt\Biggl\{
{1\over 6}\tilde\zell^{(ij)}\Phi_0\Bigl[-3\tilde Z_t^{(1)}p^2(1+
\uell^2\dell^2)+\lambda\Phi_0\uell\dell(1-2\uell\dell)\Bigr] \cr
&
-\delta^{ij}\Bigl[{1\over 4}{({\tilde Z_t^{(1)}})}^2p^2
(2+\uell^2\dell^2)-{\lambda\over 3}\Phi_0\Bigl({\lambda\over 3}\Phi_0
\tz_{\ell}\uell\dell^2-{\tz^{(1)}_t\over 2}(2-\uell\dell+2\uell^2\dell^2)
\Bigr)\Bigr]\cr
&
-{1\over 4}\tilde\zell^{(ik)}\tilde\zell^{(kj)}\Phi_0^2p^2\uell^2\dell^2
\Biggr\} .\cr}}}
In arriving at the above expressions, $O(4)$ invariance has been used:
\eqn\ofour{ \int_x \tphi\partial_{\mu}\partial_{\nu}\tphi\int_p p_{\mu}
p_{\nu} = {1\over 4}\int_x \tphi\bx\tphi\int_p p^2 .}

The coefficients obtained above become much simpler when $O(N)$
symmetry is invoked which allows us to make the following
substitutions when seeking for the RG flow equations:
\eqn\ukk{\cases{\eqalign{ \uell &~\longrightarrow U_k^{(11)}
=2(\udot+2\uddot\Phi^2) \cr
\ut &~\longrightarrow U_k^{(22)}=2\udot \cr
\lambda\Phi &~\longrightarrow U_k^{(111)}=4(3\uddot+2\udddot\Phi^2)
\vec\Phi \cr
{\lambda\over 3}\Phi &~\longrightarrow U_k^{(221)}=4\uddot\vec\Phi, \cr}}}
and
\eqn\gfd{\cases{\eqalign{ \tz^{(1)} &~\longrightarrow{\tilde{\cal Z}}_k^{(1)}
=2{{\tilde{\cal Z}}}^{'}_k\vec\Phi \cr
\tz^{(i)} &~\longrightarrow{\tilde{\cal Z}}_k^{(i)}=0 \cr
\tz^{(11)} &~\longrightarrow{\tilde{\calz}}_k^{(11)}=
4({{\tilde{\cal Z}}}^{''}_k\Phi^2+{{\tilde{\cal Z}}}^{'}_k) \cr
\tz^{(ij)} &~\longrightarrow{\tilde{\cal Z}}_k^{(ij)}=
2{{\tilde{\cal Z}}}^{'}_k\delta^{ij}, \cr}}}
where the prime notation denotes differentiation with respect to
$\Phi^2$ and we have chosen the transverse component to be along the
$2$-direction. Thus we have:
\eqn\rotr{\cases{\eqalign{ b_{k,\ell}&=-6\bigl(\tc''_{k,\ell}\Phi^2
+\tc'_{k,\ell}\bigr)\dta_{k,\ell}-\bigl(N-1\bigr)\tc'_{k,\ell}\dta_{k,t}\cr
b_{k,t}&=-2\bigl(\tc''_{k,t}\Phi^2
+\tc'_{k,t}\bigr)\dta_{k,\ell}-\bigl(N+1\bigr)\tc'_{k,t}\dta_{k,t}\cr}}}
\eqn\rro{\eqalign{ a_{k,\ell}=-4\dellk^2\Phi^2 &\Biggl\{{(\tc'_{k,\ell})}^2
k^2\Bigl[1+4\dellk^2\bigl(U_k'+2U_k''\Phi^2\bigr)^2\Bigr]+2\bigl(3U_k''
+2U_k'''\Phi^2\bigr) \cr
&
\times\biggl(\tilde{\cal Z}^{'}_{k,\ell}\Bigl[1-2(U_k'+2U_k''\Phi^2)\dellk
+8(U_k'+2U_k''\Phi^2)^2\dellk^2\Bigr] \cr
&
-4\tc_{k,\ell}\dellk^2\bigl(U_k'+2U_k''\Phi^2\bigr)\bigl(3U''_k+2U_k'''
\Phi^2\bigr)\biggr) \Biggr\},}}
\eqn\rtr{\eqalign{ a'_{k,\ell}&=4(N-1)\deltk^2\Phi^2\Biggl\{
-\tc'_{k,\ell}\bigl(\tc'_{k,t}k^2+2U_k''\bigr) \cr
&
+4U_k'\deltk\Bigl[\deltk\Bigl(2\tilde{\cal Z}_{k,t}(U_k'')^2-
(\tilde{\cal Z}_{k,t}')^2k^2U_k'\Bigr)+\tilde{\cal Z}_{k,t}'U_k''
(1-4U_k'\deltk)\Bigr]\Biggr\}.}}
\eqn\opi{\eqalign{\tilde a_{k,t}&= 2\dellk\deltk\Phi^2\Biggl\{
16\tc_{k,t}U_k'(U_k'')^2\deltk^2
-\tilde{\cal Z}'_{k,\ell}k^2\Bigl[\tc'_{k,t}+2(U_k')^2\deltk^2\bigl(
\tc'_{k,\ell}+2\tc'_{k,t}\bigr)\Bigr] \cr
&
-(\tc'_{k,t})^2k^2\Bigl(1+2(U_k')^2\deltk^2\Bigr)
+4\tc'_{k,\ell}\deltk U_k'U_k''\bigl(1-4U_k'\deltk\bigr) \cr
&
-4\tc'_{k,t}U_k''\Bigl[1-U_k'\deltk+4(U_k')^2\deltk^2\Bigr]\Biggr\},}}
\eqn\rroo{\eqalign{ a^{*}_{k,t}&=2\dellk\deltk\Phi^2\Biggl\{4U_k''\biggl[
4\tc_{k,\ell}U_k''\bigl(U_k'+2U_k''\Phi^2\bigr)\dellk^2 \cr
&
-\tc'_{k,t}\Bigl(1-(U_k'+2U_k''\Phi^2)\dellk+4(U_k'+2U_k''\Phi^2)^2\dellk^2
\Bigr)\biggr] \cr
&
+4\tc'_{k,\ell}U_k''\dellk\bigl(U_k'+2U_k''\Phi^2\bigr)\Bigl[1-4(U_k'
+2U_k''\Phi^2)\dellk\Bigr] \cr
&
-\tc'_{k,\ell}k^2\biggl[\tc'_{k,t}+2\bigl(\tc'_{k,\ell}+2\tc'_{k,t}\bigr)
\bigl(U_k'+2U''_k\Phi^2\bigr)^2\dellk^2\biggr] \Biggr\},}}
and
\eqn\triv{ a_{k,t}=a'_{k,t}=\tilde a_{k,\ell}=a^{*}_{k,\ell}=0 .}
\medskip
\bigskip
\medskip
\centerline{\bf REFERENCES}
\medskip
\nobreak
\hang\par\noindent{\wilsko} K. Wilson and J.
Kogut, {\it Phys. Rep.} {\bf 12C} (1975) 75.
\medskip
\hang\par\noindent{\decoupl} T. Appelquist and J. Carazzone, {\it Phys. Rev.}
{\bf D11} (1975) 2856.
\medskip
\par\hang\noindent{\paris} J. Polonyi, in the Proceedings of the Workshop on
``QCD Vacuum Structure and Its Applications'', edited by H.M. Fried and
B. M\"uller (World Scientific, Singapore, 1993), p. 3.
\medskip
\hang\par\noindent{\wegner} F.J. Wegner and A. Houghton, {\it Phys. Rev.}
{\bf A8} (1972) 401.
\medskip
\par\hang\noindent{\wehar}
J. Polchinski, {\it Nucl. Phys.} {\bf B231} (1984) 269;
A. Hasenfratz and P. Hasenfratz, {\it ibid.} {\bf B270} (1986) 687;
C. Wetterich, {\it ibid.} {\bf B352} (1990) 529;
N. Tetradis and C. Wetterich, {\it ibid.} {\bf B398} (1993) 659;
P. Hasenfratz and J. Nager, {\it Z. Phys.} {\bf C37} (1988) 477;
A. Margaritis, G. Odor and A. Patkos, {\it ibid.} {\bf C39} (1988) 109;
M. Alford, {\it Phys. Lett.} {\bf B336} (1994) 237;
T. R. Morris, {\it ibid.} {\bf B329} 241, {\bf B334} (1994) 355;
R. D. Ball and R. S. Thorne, {\it Ann. Phys.} {\bf 236} (1994) 117.
\medskip
\hang\par\noindent{\lp} S.-B. Liao and J. Polonyi, {\it Ann. Phys.}
{\bf 222} (1993) 122.
\medskip
\hang\par\noindent{\fraser} C. M. Fraser, {\it Z. Phys.} {\bf C28} (1985) 101.
\medskip
\par\hang\noindent{\patr} A. Patrascioiu, J.L. Richard and E. Seiler,
{\it Phys. Lett.} {\bf B254} (1991) 173.
\medskip
\par\hang\noindent{\itzykson} C. Itzykson and J. Zuber, {\it Quantum
Field Theory} (McGraw-Hill, New York, 1980), p.204.
\vfill
\end